\documentclass[12pt]{amsart}
\usepackage{amsfonts,amsthm,amssymb,amscd}
\usepackage[mathscr]{eucal}
\usepackage{stmaryrd}
\usepackage{graphicx}
\usepackage[curve,matrix,arrow]{xy}

\advance\textheight\topskip
\usepackage[bookmarks=false, a4paper, hyperfigures=false]{hyperref}
\usepackage{times}
\allowdisplaybreaks

\numberwithin{equation}{section}
\makeatletter
\@addtoreset{equation}{section}
\@addtoreset{subsubsection}{section}

\def\@secnumfont{\bfseries}
\def\subsubsection{\@startsection{subsubsection}{3}%
  \z@{.5\linespacing\@plus.7\linespacing}{-.5em}%
  {\normalfont\bfseries}}
\def\paragraph{\@startsection{paragraph}{4}%
  \z@\z@{-\fontdimen2\font}%
  \normalfont\bfseries}
\def\subparagraph{\@startsection{subparagraph}{5}%
  \z@\z@{-\fontdimen2\font}%
  \normalfont\bfseries}

\makeatother

\newcommand{\soc}{\mathop{\mathrm{soc}}\nolimits}

\newcommand{\Vir}{\mathrm{Vir}^{\phantom{y}}_{\kern-1pt\frac{p_+}{p_-}}}

\newcommand{\seti}{\mathscr{I}_0}
\newcommand{\setii}{\mathscr{I}_1}
\newcommand{\setp}{\mathscr{I}_{\boxslash}}
\newcommand{\setm}{\mathscr{I}_{\boxbslash}}

\newcommand{\wkappap}{\varkappa}
\newcommand{\wkappam}{\chi}

\newcommand{\wrhop}{\rho^{\boxslash}}

\newcommand{\wrhom}{\rho^{\boxbslash}}

\newcommand{\wphip}{\varphi^{\boxslash}}
\newcommand{\wphim}{\varphi^{\boxbslash}}

\newcommand{\wxi}{\rho}
\newcommand{\wpsi}{\psi}
\newcommand{\wzeta}{\varphi}

\newcommand{\kk}{k}
\newcommand{\ep}{e_+}
\newcommand{\fp}{f_+}
\newcommand{\emi}{e_-}
\newcommand{\fm}{f_-}
\newcommand{\epm}{e_\pm}
\newcommand{\fpm}{f_\pm}
\newcommand{\qp}{\mathfrak{q}_+}
\newcommand{\qm}{\mathfrak{q}_-}
\newcommand{\qpm}{\mathfrak{q}_\pm}
\newcommand{\dkk}{\kappa}

\newcommand{\qalgA}{\boldsymbol{\mathfrak{g}}} 
\newcommand{\tqalgA}{\qalgA_{p_+,p_-}}
\newcommand{\XXX}{\qalgA_{p_+,p_-}}
\newcommand{\qalgB}{\mathscr{H}}   

\newcommand{\voal}[1]{\boldsymbol{\mathscr{#1}}} 
\newcommand{\algW}{\voal{W}}
\newcommand{\talgW}{\voal{W}_{p_+,p_-}}
\newcommand{\WWW}{\voal{W}_{p_+,p_-}}

\newcommand{\talgM}{\voal{M}_{p_+,p_-}}

\newcommand{\algL}{\voal{L}}
\newcommand{\talgL}{\voal{L}_{p_+,p_-}}

\newcommand{\algR}{\voal{R}}

\newcommand{\rep}{\mathcal}  
\newcommand{\trep}{\mathscr} 

\newcommand{\repX}{\rep{X}}
\newcommand{\repV}{\rep{V}}
\newcommand{\repP}{\rep{P}}

\newcommand{\repK}{\rep{K}}

\newcommand{\tvacL}{\trep{L}_{p_+,p_-}}

\newcommand{\tvacW}{\trep{W}_{p_+,p_-}}

\newcommand{\repJ}{\rep{J}}

\newcommand{\repF}{\rep{F}} 

\newcommand{\bigrepF}{\mathbb{F}}

\newcommand{\XX}{\mathsf{X}} 

\newcommand{\cat}{\mathfrak}

\newcommand{\catW}{\cat{W}}

\newcommand{\TA}{\mathscr{C}} 
\newcommand{\sgch}{{\bar{\mathscr{C}}}} 

\newcommand{\cZcft}{\mathfrak{Z}_{\mathrm{cft}}}

\newcommand{\Grring}{\mathscr{G}}

\newcommand{\polP}{\mathcal{P}}

\newcommand{\scrBox}{{\scriptstyle\Box}}
\newcommand{\End}{\mathrm{End}}
\newcommand{\im}{\mathop{\mathrm{im}}\nolimits}
\renewcommand{\ker}{\mathop{\mathrm{ker}}\nolimits}
\renewcommand{\geq}{\,{\geqslant}\,}
\renewcommand{\leq}{\,{\leqslant}\,}
\newcommand{\tensor}{\otimes}

\newcommand{\half}{%
  \mathchoice{\ffrac{1}{2}}{\frac{1}{2}}{\frac{1}{2}}{\frac{1}{2}}}
\newcommand{\dd}{\partial}
\newcommand{\one}{\boldsymbol{1}}
\newcommand{\Tr}{\mathrm{Tr}^{\vphantom{y}}}
\newcommand{\pbw}{\boldsymbol{e}}
\newcommand{\coup}[2]{\langle#1,#2\rangle} 

\newcommand{\SLiiZ}{SL(2,\oZ)}

\newcommand{\modS}{\mathscr{S}}
\newcommand{\modT}{\mathscr{T}}

\newcommand{\repLy}{\pi}
\newcommand{\repSw}{\bar\pi}
\newcommand{\repA}{\pi^*}

\newcommand{\oC}{\mathbb{C}}
\newcommand{\oN}{\mathbb{N}}
\newcommand{\oZ}{\mathbb{Z}}

\newcommand{\bref}[1]{\textbf{\ref{#1}}}
\newcommand{\ffrac}[2]{\mbox{\footnotesize$\displaystyle\frac{#1}{#2}$}}
\newcommand{\sfrac}[2]{\mbox{\scriptsize$\displaystyle\frac{#1}{#2}$}}
\renewcommand{\tilde}{\widetilde}
\newcommand{\captionfont}[1]{\textit{\textbf{\footnotesize#1}}} 

\newcommand{\q}{\mathfrak{q}}
\newcommand{\qint}[1]{[#1]}
\newcommand{\qbin}[2]{\mathchoice%
  {{\qbinm{#1}{#2}}}{\qbinmm{#1}{#2}}%
  {\qbinmm{#1}{#2}}{\qbinmm{#1}{#2}}}
\newcommand{\qbinm}[2]{\mbox{\footnotesize$\displaystyle
    \genfrac{[}{]}{0pt}{}{#1}{#2}$}}
\newcommand{\qbinmm}[2]{\genfrac{[}{]}{0pt}{}{#1}{#2}}

\swapnumbers
\newtheorem{Thm}[subsection]{Theorem}

\newtheorem{lemma}[subsubsection]{Lemma}

\newtheorem{prop}[subsubsection]{Proposition}

\newtheorem{Conj}[subsection]{Conjecture}

\theoremstyle{definition}
\newtheorem{Dfn}[subsection]{Definition}
\newtheorem{dfn}[subsubsection]{Definition}
\newtheorem{rem}[subsubsection]{Remark}

\begin{document}

\title[Logarithmic extensions of minimal models]{%
  \vspace*{-4\baselineskip}
  \mbox{}\hfill\texttt{\small\lowercase{hep-th}/\lowercase{0606196}}
  \\[\baselineskip]
  Logarithmic extensions of minimal models: characters and modular
  transformations}

\author[Feigin]{B.L.~Feigin}%

\address{\mbox{}\kern-\parindent blf: Landau Institute for Theoretical
  Physics \hfill\mbox{}\linebreak \texttt{feigin@mccme.ru}}

\author[Gainutdinov]{A.M.~Gainutdinov}%

\address{\mbox{}\kern-\parindent ams, amg, iyt: Lebedev Physics
  Institute \hfill\mbox{}\linebreak \texttt{ams@sci.lebedev.ru},
  \texttt{azot@mccme.ru}, \texttt{tipunin@td.lpi.ru}}

\author[Semikhatov]{A.M.~Semikhatov}%

\author[Tipunin]{I.Yu.~Tipunin}

\begin{abstract}
  We study logarithmic conformal field models that extend the $(p,q)$
  Virasoro minimal models.  For coprime positive integers $p$ and $q$,
  the model is defined as the kernel of the two minimal-model
  screening operators.  We identify the field content, construct the
  $W$-algebra $\algW_{p,q}$ that is the model symmetry (the maximal
  local algebra in the kernel), describe its irreducible modules, and
  find their characters.  We then derive the $\SLiiZ$-representation
  on the space of torus amplitudes and study its properties.  {}From
  the action of the screenings, we also identify the quantum group
  that is Kazhdan--Lusztig-dual to the logarithmic model.
\end{abstract}

\maketitle

\thispagestyle{empty}

\setcounter{tocdepth}{2}

\vspace*{-24pt}

\begin{footnotesize}\addtolength{\baselineskip}{-6pt}
  \tableofcontents
\end{footnotesize}

\section{Introduction}

\noindent\textbf{1.1.}\ \
\addtocounter{subsection}{1}Logarithmic conformal field theory can be
interesting for two reasons at least.  The first is their possible
applications in condensed-matter systems: the quantum Hall
effect~\cite{[GurarieFlohr],[Saleur1],[Saleur2]}, self-organized
critical phenomena~\cite{[Ruelle],[Jeng]}, the two-dimensional
percolation problem~\cite{[Cardy],[Watts],[DuplantierSaleur]}, and
others (see, e.g.,~\cite{[GL]} and the references therein).  The
second is the general category-theory aspects of conformal field
theory involving vertex-operator algebras with nonsemisimple
representation categories~\cite{[HLZ],[Mi]} (also see~\cite{[Fuchs]}
and the references therein).  Unfortunately, there are few
well-investigated examples in logarithmic conformal field theory.
Presently, only the $(2,1)$
model~\cite{[K-first],[GaberdielKausch1],[GaberdielKausch3],
  [Kausch-sympl],[FFHST],[FGST2]} is formulated with sufficient
completeness.  A number of results exist on the $(p,1)$
models~\cite{[Flohr],[GaberdielKausch],[FHST],[FGST]}, but no model
has been investigated so completely as the Ising model, for example.

An essential new feature of nonsemisimple (logarithmic) conformal
field theories, in comparison with semisimple (rational) theories,
already occurs in constructing the space of states.  In the semisimple
case, it suffices to take the sum of all irreducible representations
in each chiral sector.  But in the nonsemisimple case, there are
various indecomposable representations, constructed beginning with
first extensions of irreducible representations, then taking their
further extensions, and so on, ending up with \textit{projective
  modules} (the largest possible indecomposable extensions, and hence
the modules with the largest Jordan cells, for the scaling dimension
operator $L_0$, that can be constructed for a given set of irreducible
representations).  The space of states is therefore given by the sum
over all nonisomorphic indecomposable projective modules,
\begin{equation}\label{proj-sum}
  \mathbb{P}=\bigoplus_\iota \rep{P}_\iota.
\end{equation}
This affects not only theory but also applications: the physically
important operators (thermal, magnetic, and so on) in specific models
may often be identified with the field corresponding to the
``highest-weight'' vector in a projective module $\rep{P}_\iota$, not
with the primary field (of the same dimension) corresponding to the
highest-weight vector of the irreducible quotient of~$\rep{P}_\iota$.
Also, whenever indecomposable representations are involved, there are
more possibilities for constructing modular invariants by combining
the chiral and antichiral spaces of states.

But an even more essential point about nonsemisimple theories is that
before speaking of the representations, their characters, fusion,
etc., one must find the symmetry algebra whose representations,
characters, fusion, etc.\ are to be considered; this algebra is
typically \textit{not} the ``naive,'' manifest symmetry algebra.  This
point was expounded in~\cite{[FHST]}.  The symmetry algebras of
logarithmic conformal field theory models are typically nonlinear
extensions of the naive symmetry algebra (e.g., Virasoro), i.e., are
some $W$-algebras.  The first examples of $W$-algebras arising in this
context were studied
in~\cite{[K-first],[GaberdielKausch1],[GaberdielKausch3]}, also
see~\cite{[CF]} and the references therein.  

In a given logarithmic model, the chiral $W$-algebra must be big
enough to ensure that only a finite number of its irreducible
representations are realized in the model.  Only then can one expect
to have finite-dimensional fusion rules and modular group
representation (we recall that a finite-dimensional modular group
representation is a good test of the \textit{consistency} of the
model.  Also, once finitely many irreducible representations are
involved, there are finitely many projective modules, and the sum
in~\eqref{proj-sum} makes sense.

In this paper, we systematically investigate the logarithmic
extensions of the $(p,q)$ models of conformal field theory.  This task
was already set in~\cite{[FGST]}, in the line of an appropriate
extension of the results obtained for the $(p,1)$ models.  The method
in~\cite{[FHST],[FGST]} is to define and construct logarithmic
conformal field models in terms of free fields and screenings.  The
chiral algebra~$\algW$ that is the symmetry algebra of the model is
then derived from the kernel of the screenings in the vacuum
representation of a lattice vertex-operator algebra~$\algL$.
Irreducible representations of~$\algW$ are identified with the images
and cohomology of (certain powers of) the screenings in irreducible
representations of~$\algL$.  The projective $\algW$-modules
in~\eqref{proj-sum} are then to be constructed as the projective
covers of the irreducible representations (this construction may be
rather involved; a notable exception is provided by the $(2,1)$ model,
see~\cite{[Kausch-sympl],[FFHST],[FGST2]}).

In what follows, we set $p=p_+$ and $q=p_-$, a fixed pair of coprime
positive integers.

In the \textit{rational} $(p_+,p_-)$ model, the chiral symmetry
algebra is the vertex-operator algebra~$\voal{M}_{p_+,p_-}$ defined as
the cohomology of screenings that act in the vacuum
representation~$\trep{L}_{p_+,p_-}$ of the appropriate lattice
vertex-operator algebra~$\voal{L}_{p_+,p_-}$; irreducible
representations of~$\voal{M}_{p_+,p_-}$ can then be identified with
the cohomology of (powers of) the screenings in irreducible
representations of~$\voal{L}_{p_+,p_-}$.

In the \textit{logarithmic} $(p_+,p_-)$ model, (the vacuum
representation of) the chiral algebra $\voal{W}_{p_+,p_-}$
extends~$\voal{M}_{p_+,p_-}$ such that
$\voal{M}_{p_+,p_-}=\voal{W}_{p_+,p_-}/\algR$, where~$\algR$ is the
maximal vertex-operator ideal in~$\voal{W}_{p_+,p_-}$.  The
$W$-algebra $\voal{W}_{p_+,p_-}$ can be defined as the intersection of
the screening kernels in~$\trep{L}_{p_+,p_-}$. (As we see in what
follows, $\talgW$ is generated by the energy--momentum tensor $T(z)$
and two Virasoro primaries $W^+(z)$ and $W^-(z)$ of conformal
dimension~$(2p_+{-}1)(2p_-\,{-}\,1)$.)  The irreducible
representations of $\voal{W}_{p_+,p_-}$ are of two different kinds.
The first are the $\half(p_+\,{-}\,1)(p_-\,{-}\,1)$ irreducible
modules of the Virasoro minimal model or, in other words, the modules
annihilated by~$\algR$.  The second are the $2 p_+ p_-$ modules that
admit a nontrivial action of~$\algR$. \ They can be identified with
the images of (powers of) the screenings in the respective irreducible
representations of the lattice vertex-operator algebra
$\voal{L}_{p_+,p_-}$ and decompose into infinitely many irreducible
Virasoro modules.  (The Virasoro embedding structure in the $(3,2)$
and $(5,2)$ logarithmic models was recently arduously explored
in~\cite{[Flohrpp]}.)

The characters of the $\half(p_+\,{-}\,1)(p_-\,{-}\,1) + 2 p_+ p_-$
irreducible representations of $\voal{W}_{p_+,p_-}$ (or, in slightly
different terms, the \textit{Grothendieck ring}\footnote{The free
  Abelian group generated by symbols $[M]$, where $M$ ranges over all
  representations subject to relations $[M]=[M']+ [M'']$ for all exact
  sequences $0 \to M'\to M\to M''\to 0$.} $\Grring$) do not exhaust
the space~$\TA$ of torus amplitudes of the logarithmic $(p_+,p_-)$
model: we only have that $\TA\supset\Grring$.  This is a
characteristic feature of nonsemisimple (logarithmic) conformal field
theories, cf.~\cite{[F],[KeLu],[FHST],[My3],[FG],[FGST]}. \ Because
$\TA$ carries a representation of~$\SLiiZ$, a much better
approximation to this space is $\sgch$, the $\SLiiZ$-representation
generated from~$\Grring$:
\begin{equation*}
  \TA\supset\sgch\supset\Grring
\end{equation*}
(where most probably $\TA=\sgch$).

\begin{Thm}\label{thm:R-decomp}
  In the logarithmic $(p_+,p_-)$ model,
  \begin{enumerate}
    
  \item $\dim\Grring=2p_+p_- + \half(p_+\,{-}\,1)(p_-\,{-}\,1)$;
    
  \item $\dim\sgch=\half(3p_+\,{-}\,1)(3p_-\,{-}\,1)$;
    
  \item the $\SLiiZ$-representation $\repLy$ on~$\sgch$ has the
    structure
    \begin{equation}\label{eq:sl2z-dec}
      \sgch=R_{\mathrm{min}}\oplus
      R_{\mathrm{proj}}\oplus\oC^2\tensor\bigl(
      R_{\boxslash}\oplus R_{\boxbslash}\bigr)
      \oplus\oC^3\tensor R_{\mathrm{min}},
    \end{equation}
    where $\oC^2$ is the standard two-dimensional representation,
    $\oC^3\cong\mathrm{S}^2(\oC^2)$ is its symmetrized square,
    $R_{\mathrm{min}}$ is the
    $\half(p_+\,{-}\,1)(p_-\,{-}\,1)$-dimensional
    $\SLiiZ$-represen\-ta\-tion on the characters of the $(p_+,p_-)$
    Virasoro minimal model, and $R_{\mathrm{proj}}$,
    $R_{\boxslash}$, and $ R_{\boxbslash}$ are
    $\SLiiZ$-representations of the respective dimensions
    $\half(p_+\,{+}\,1)\cdot{}$\linebreak[0]$(p_-\,{+}\,1)$,
    $\half(p_+\,{-}\,1)(p_-\,{+}\,1)$, and
    $\half(p_+\,{+}\,1)(p_-\,{-}\,1)$.
  \end{enumerate}
\end{Thm}

We note that the space $R_{\mathrm{min}}\oplus R_{\mathrm{proj}}$ is
spanned by the characters of irreducible modules of the lattice
vertex-operator algebra~$\talgL$ and coincides with the space of theta
functions.\footnote{Technically, the extension from
  $R_{\mathrm{min}}\oplus R_{\mathrm{proj}}$ to $\sgch$ involves
  derivatives of the theta functions, which gives rise to the explicit
  occurrences of the modular parameter~$\tau$ equalizing the modular
  weights.  The highest theta-function derivative and simultaneously
  the top power of $\tau$ thus occurring, $n$, is equal to~$2$ for
  $(p_+,p_-)$ models and to~$1$ for $(p,1)$ models~\cite{[FGST]},
  which can be considered the origin of the corresponding $\oC^{n+1}$
  in~\eqref{eq:sl2z-dec} and in a similar formula in~\cite{[FGST]}.}
The space $R_{\mathrm{proj}}$ can be identified with the characters of
projective $\talgW$-modules.

\begin{rem}
  The $2p_+p_- + \half(p_+\,{-}\,1)(p_-\,{-}\,1)$ irreducible
  $\WWW$-representations are rather naturally arranged into a Kac
  table as follows.  First, the $\half(p_+\,{-}\,1)(p_-\,{-}\,1)$
  rational-model representations occupy the standard positions, with
  the standard identifications, in the boxes of the standard
  $(p_+\,{-}\,1)\times(p_-\,{-}\,1)$ Kac table.  Next, each box of the
  \textit{extended} $p_+\times p_-$ Kac table contains two, a ``plus''
  and a ``minus,'' of the $\WWW$-representations labeled
  $(\repX^\pm_{r,s})_{\substack{1\leq r\leq p_+\\ 1\leq s\leq p_-}}$
  in what follows.  We note that a logarithmic conformal field theory
  containing only finitely many irreducible or indecomposable
  \textit{Virasoro} representations does not seem to exist; in
  particular, each $(p_+,p_-)$ model considered in this paper contains
  infinitely many indecomposable Virasoro representations, usually
  with multiplicity greater than~$1$, and we use the term ``Kac
  table'' exclusively to refer to a finite set of irreducible
  representations of the $\talgW$ algebra.
\end{rem}

Theorem~\bref{thm:R-decomp} also implies that there exist
$\SLiiZ$-representations $\repSw$ and $\repA$ on~$\sgch$ such that
\begin{equation*}
  \repLy(\gamma)=\repA(\gamma)\repSw(\gamma),\quad
  \repSw(\gamma)\repA(\gamma') =\repA(\gamma')\repSw(\gamma),
  \qquad
  \gamma,\gamma' \in\SLiiZ.
\end{equation*}
The representation $\repSw$ can be restricted to $\Grring$, which then
decomposes in terms of $\SLiiZ$ representations as
\begin{equation*}
  \Grring=R_{\mathrm{min}}\oplus
  R_{\mathrm{proj}}\oplus
  R_{\boxslash}\oplus R_{\boxbslash}
  \oplus R_{\mathrm{min}}.
\end{equation*}
This decomposition can be taken as the starting point for constructing
a logarithmic Verlinde formula as in~\cite{[FHST]}.

In view of the fundamental importance of the $\SLiiZ$ action, this
theorem gives a very strong indication regarding the field content of
a consistent conformal field theory model: the
$\half(3p_+\,{-}\,1)(3p_-\,{-}\,1)$-dimensional space of
\textit{generalized characters} $\sgch$ is a strong candidate for the
space of torus amplitudes of the logarithmic $(p_+,p_-)$ model.  The
following conjecture appears to be highly probable.
\begin{Conj} \label{conj:gch}
  The $\SLiiZ$-representation generated from~$\Grring$ coincides with
  the space of torus amplitudes:\\[-10pt]
  \begin{equation*}
    \sgch=\TA
  \end{equation*}
  as $\SLiiZ$ representations.  
\end{Conj}

\medskip

\noindent\textbf{1.4.}\ \ \
\addtocounter{subsection}{1}To move further in constructing the full
space of states of the logarithmic theory as in~\eqref{proj-sum}, one
must construct projective covers of all the $2p_+p_- +
\half(p_+\,{-}\,1)(p_-\,{-}\,1)$ irreducible $\WWW$-representations.
This is a separate, quite interesting task.\footnote{In particular,
  the question about logarithmic partners of the energy--momentum
  tensor $T(z)$ and other fields takes the form of a well-posed
  mathematical problem about the structure of projective
  $\WWW$-modules.  Logarithmic partners have been discussed, e.g.,
  in~\cite{[GL],[GL-2002],[KNi],[KNi2],[FMl]} in the case $c=0$, where
  the differential polynomial in~$T(z)$ whose logarithmic partner is
  sought coincides with~$T(z)$ itself.}

Some useful information on the structure of the chiral-algebra
projective modules can be obtained from the Kazhdan--Lusztig
correspondence.  In general, it is a correspondence between a chiral
algebra $\algW$ and its representation category~$\catW$ realized in
conformal field theory, on the one hand, and some ``dual'' quantum
group and its representation category on the other hand.  In some
``well-behaved'' cases, the occurrence of the Kazhdan--Lusztig-dual
quantum group can be seen by taking $\bigoplus_\iota
d_\iota\rep{P}_\iota$, a direct sum of projective $\algW$-modules with
some multiplicities~$d_\iota$ chosen such that they are additive with
respect to the direct sum\footnote{For a given chiral algebra $\algW$,
  such sums are assumed to be finite; from a somewhat more general
  standpoint, this must follow from a set of fundamental requirements
  on~$\algW$ in a given nonsemisimple model.  First, the algebra
  itself must be generated by a finite number of fields~$W_i(z)$.  The
  category~$\catW$ of $\algW$-modules with locally nilpotent action of
  the positive modes of~$W_i(z)$ is then well defined.  Second, the
  $\algW$-modules must be finitely generated, such that for any
  collection of positive integers $(N_i)$, the coinvariants with
  respect to the subalgebra $\algW(N_1,\dots,N_m)\subset\algW$
  generated by the modes $W_i[n_i]$ with $n_i\leq N_i$ be
  finite-dimensional in any module from~$\catW$.  In particular, this
  means that the category~$\catW$ contains a finite number of
  irreducible representations, and hence a finite number of projective
  modules (cf.~\cite{[DLM],[My3]}).} and multiplicative with respect
to the quasitensor product in~$\catW$.  If such a choice of
the~$d_\iota$ is possible, then
\begin{equation}
  \qalgA=\End_{\algW}\Bigl(\bigoplus_\iota d_\iota\rep{P}_\iota\Bigr)
\end{equation}
can be endowed with the structure of a (Hopf) algebra and $d_\iota$
are the dimensions of its irreducible representations (constructing
the comultiplication requires certain conditions, which we do not
discuss here).  In this case, the category $\catW$ is equivalent (as a
quasitensor category) to the category of finite-dimensional
representations of~$\qalgA$.  \ Such an extremely well-behaved case is
realized in $(p,1)$ logarithmic models~\cite{[FGST2]}: classification
of indecomposable representation of the Kazhdan--Lusztig-dual quantum
group $\qalgA$, which is not difficult to obtain using quite standard
means, gives the classification of indecomposable
$\voal{W}_{p,1}$-representations.
In particular, the structure of projective $\voal{W}_{p,1}$-modules is
thus known.

In the $(p_+,p_-)$ logarithmic models, the Kazhdan--Lusztig-dual
quantum group $\XXX$ (obtained as a subalgebra in the quotient of the
Drinfeld double of the algebra of screenings for the $\WWW$ algebra)
is not Morita-equivalent to $\WWW$, but nevertheless provides
important information on the structure of indecomposable $\WWW$
modules.  On the one hand, the quantum group $\XXX$ and its
representation category give only an ``approximation'' to the
structure of the $\WWW$-representation category (the representation
categories are certainly not equivalent, in contrast to the $(p,1)$
case~\cite{[FGST2]}; in particular, there are $2p_+ p_-$
indecomposable projective $\XXX$-modules but $2p_+ p_- +
\half(p_+\,{-}\,1)(p_-\,{-}\,1)$ indecomposable projective
$\WWW$-modules).  On the other hand, this ``approximation'' becomes
the precise correspondence as regards the modular group
representations: naturally associated with $\XXX$ is the
$\SLiiZ$-representation on its center~\cite{[FGST-q]}, which turns out
to be \textit{equivalent} to the $\SLiiZ$-representation on the
$\WWW$-algebra characters and generalized characters
in~\eqref{eq:sl2z-dec}.

The $\XXX$ quantum group acts in the space~$\bigrepF$ obtained as a
certain extension (``dressing'') of the irreducible
$\talgL$-modules.  Moreover, $\bigrepF$ is in fact a
$(\talgW,\tqalgA)$-bimodule.  This bimodule structure plays an
essential role in the description of the full conformal field theory
on Riemann surfaces of different genera, defect lines, boundary
conditions, etc.

\medskip

\noindent\textbf{1.5.}\ \ \
\addtocounter{subsection}{1}{}From the standpoint of applications in
condensed-mater physics, the definition of both the $W$-algebra $\WWW$
and the quantum group $\XXX$ refers to the Coulomb-gas
picture~\cite{[Nien],[DotsFat]}, where the starting point is a
two-dimensional scalar field $\varphi$ with the action written in
complex coordinates as
\begin{equation*}
  S_0=-\ffrac{1}{8\pi}\int\dd\varphi\bar\dd\varphi dzd\bar z.
\end{equation*}
(The normalization is chosen such that the propagator has the form
$\langle\varphi(z,\bar z)\varphi(0,0)\rangle=\log|z|$ and vertex
operators $\exp(\alpha\varphi(z,\bar z))$ do not involve an $i$ in the
exponent.)  Furthermore, the field is taken to be compactified to a
circle (of the radius~$\sqrt{\mathstrut 2p_+p_-}$), which just means
that the fields $\varphi$ and $\varphi+2i\pi\sqrt{\mathstrut 2p_+p_-}$
are considered equivalent.  Then the observables are given by vertex
operators $\exp(\frac{n}{\sqrt{\mathstrut 2p_+p_-}}\varphi(z,\bar z))$
with $n\in\oZ$.

This model with central charge $c=1$ has a large symmetry algebra, the
lattice vertex-operator algebra $\talgL$ mentioned above.  The minimal
models can be regarded as conformal points with $c<1$~\cite{[Zamolod]}
to which the system renormalizes after the perturbation
\begin{equation*}
  S=S_0+\int\!\lambda_+ e^{\alpha_+\varphi(z,\bar z)}dzd\bar z+
  \int\!\lambda_- e^{\alpha_-\varphi(z,\bar z)}dzd\bar z
\end{equation*}
with the appropriate $\alpha_+$ and $\alpha_-$ (and some constant
$\lambda_\pm$).  The symmetry of the model thus obtained is the
vertex-operator algebra $\talgM$, which is ``much smaller''
than~$\talgL$.

Logarithmic conformal points occur in this approach for quenched
random systems~\cite{[qCardy]} whose action is given by
\begin{equation*}
  S=S_0+\int \lambda_+(z,\bar z) e^{\alpha_+\varphi(z,\bar z)}dzd\bar z+
  \int \lambda_-(z,\bar z) e^{\alpha_-\varphi(z,\bar z)}dzd\bar z,
\end{equation*}
where $\lambda_\pm(z,\bar z)$ are quenched random variables with
appropriately chosen correlators $\overline{\lambda_\pm(z,\bar z)}$,
$\overline{\lambda_\pm(z_1,\bar z_1)\lambda_\pm(z_2,\bar z_2)}$, and
so on.  The parameters involved in these fixed correlators renormalize
such that the system occurs in a new infrared fixed point with the
same $c<1$ as in the minimal model, but with the symmetry algebra
given by a $W$-algebra (a subalgebra of $\talgL$).  The entire system
can thus be regarded as the tensor product of the original Coulomb-gas
model and an additional model describing a quenched disorder through
the chosen correlators of the $\lambda_\pm(z,\bar z)$.  We do not know
how these correlators must be chosen in order to produce just the
$(p_+,p_-)$ logarithmic conformal field theory model; instead, we take
a more algebraic stand and study the $W$-algebra $\WWW$ of the
expected fixed point.  It is then possible to make contact with
quenched disorder by studying the projective modules of this
$W$-algebra.  

\medskip

This paper is organized as follows.  In Sec.~\ref{sec:free-field}, we
introduce the basic notation and describe some facts in the free-field
description of minimal models.  We introduce vertex operators
in~\bref{subsec:vertex}, define the free-field and Virasoro modules
in~\bref{subsec:modules}, and introduce the lattice vertex-operator
algebra~$\talgL$ and its modules that are important ingredients in
constructing the $\WWW$-representations (\bref{sec:lvoa}).  In
Sec.~\ref{sec:screenings}, we describe the action of screenings on the
vertices in~\bref{subsec:screenings} and introduce the
Kazhdan--Lusztig-dual quantum group~$\XXX$ (we reformulate the action
of the screenings in terms of a Hopf algebra $\qalgB$
in~\bref{subsec:q-gr}, and then construct its Drinfeld
double~$\tqalgA$ in~\bref{thm:double}).\ \ In Sec.~\ref{sec:w-rep}, we
finally construct the $\talgW$ algebra (we introduce it
in~\bref{subsec:w-def}, formulate the structural result
in~\bref{prop:w-alg}, and describe the irreducible and Verma
$\talgW$-modules in~\bref{subsec:w-mod} and~\bref{subsec:Verma}; a
$(\talgW,\tqalgA)$-bimodule structure of the space of states in
given~\bref{subsec:bimodule}).  In Sec.~\ref{sec:torus}, we calculate
the $\SLiiZ$-representation generated from the $\WWW$-characters; on
the resulting space of generalized characters (as noted above, most
probably the torus amplitudes), we then decompose the $\SLiiZ$-action
as in~\eqref{eq:sl2z-dec}.  In~\bref{subsec:char}, we calculate
characters of the irreducible $\talgW$-modules; in~\bref{subsec:gch},
we introduce generalized characters, calculate the~$\SLiiZ$ action
on~$\sgch$, and give a direct proof~of~\bref{thm:R-decomp}.  Several
series of modular invariants are considered in~\bref{sec:mod-inv}.
Some implications of the Kazhdan--Lusztig correspondence and several
open problems are mentioned in the conclusions.

\subsection*{Notation} 
We fix two coprime positive integers $p_+$ and $p_-$ and set
\begin{equation}\label{eq:numbers}
  \alpha_-=-\sqrt{\ffrac{2p_+}{p_-}},
  \quad
  \alpha_+=\sqrt{\ffrac{2p_-}{p_+}},\quad
  \alpha_0=\alpha_++\alpha_-.
\end{equation}
With the pair $(p_+, p_-)$, we associate the sets of indices
\begin{align}
  \seti&=\Bigl\{(r,s)\,\,|\,\, 0\leq r\leq p_+,\,\, 0\leq s\leq
  p_-,\,\,
  p_- r+ p_+s\leq p_+p_-,\,\, (r,s)\neq(0,p_-)\Bigr\},\\
  \label{eq:setii}
  \setii&=\Bigl\{(r,s)\,\,|\,\, 1\leq r\leq p_+\!-\!1,\quad 1\leq
  s\leq p_-\!-\!1,\quad
  p_- r+ p_+s\leq p_+p_-\Bigr\},\\
  \setm&=\Bigl\{(r,s)\,\,|\,\, 0\leq r\leq p_+\!-\!1,\quad 1\leq s\leq
  p_-\!-\!1,\quad
  p_- r+ p_+s\leq p_+p_-\Bigr\},\\
  \setp&=\Bigl\{(r,s)\,\,|\,\, 1\leq r\leq p_+\!-\!1,\quad 0\leq s\leq
  p_-\!-\!1,\quad p_- r+ p_+s\leq p_+p_-\Bigr\}.
\end{align}
The numbers of elements in these sets are
$|\seti|=\half(p_+\,{+}\,1)(p_-\,{+}\,1)$,
$|\setii|=\half(p_+\,{-}\,1)\cdot{}$\linebreak[0]$(p_-\,{-}\,1)$,
$|\setm|=\half(p_+\,{+}\,1)(p_-\,{-}\,1)$, and
$|\setp|=\half(p_+\,{-}\,1)(p_-\,{+}\,1)$.

The building blocks for the characters are the so-called
theta-constants $\theta_{s,p}(q)$, $\theta^{(1)}_{s,p}(q)$, and
$\theta^{(2)}_{s,p}(q)$, where
\begin{equation}\label{theta-const}
  \theta_{s,p}(q)=\theta_{s,p}(q,1),\qquad
  \theta^{(m)}_{s,p}(q)=\Bigl(z\ffrac{\dd}{\dd z}\Bigr)^{\!m}
  \theta_{s,p}(q,z)\Bigr|_{z=1},\quad m\in\oN,
\end{equation}
and the theta function is defined as
\begin{equation*}
  \theta_{s,p}(q,z)=\sum_{j\in\oZ + \frac{s}{2p}} q^{p j^2} z^{p j},
  \quad |q|<1,~z\in\oC,\quad p\in\oN,\quad s\in\oZ.
\end{equation*}

To further simplify the notation, we resort to the standard abuse by
writing $f(\tau)$ for $f(e^{2i\pi\tau})$, with $\tau$ in the upper
complex half-plane; it is tacitly assumed that $q=e^{2i\pi\tau}$.

There are the easily verified properties
\begin{equation*}
  \theta^{(m)}_{s+2pa,p}(\tau)=\theta^{(m)}_{s,p}(\tau),\qquad
  \theta^{(m)}_{-s,p}(\tau)=(-1)^m\theta^{(m)}_{s,p}(\tau),
  \quad p\in\oN,\quad m\in\oN_0,\quad a\in\oZ
\end{equation*}
and $\theta'_{0,p}(\tau)=\theta'_{p,p}(\tau)=0$.
We often write
\begin{equation*}
 \theta_{s,p_+p_-}(\tau)\equiv\theta_{s},
  \qquad\theta^{(1)}_{s,p_+p_-}(\tau)\equiv\theta'_{s},
  \qquad\theta^{(2)}_{s,p_+p_-}(\tau)\equiv\theta''_{s}.
\end{equation*}

Similar abbreviations are used for the characters: we write
\begin{equation*}
  \chi_{r,s}(\tau)\equiv\chi_{r,s},
  \qquad\chi^{\pm}_{r,s}(\tau)\equiv\chi^{\pm}_{r,s}.
\end{equation*}

Finally, we use the $\eta$-function
\begin{equation*}
  \eta(q)=q^{\frac{1}{24}} \prod_{n=1}^{\infty} (1-q^n).
\end{equation*}

\section{Free-field preliminaries}\label{sec:free-field}
We introduce a free bosonic field and describe its energy--momentum
tensor, vertex operators, screenings, and representations of the
Virasoro algebra.

\subsection{Free field and vertices}\label{subsec:vertex}
Let $\varphi$ denote a free scalar field with the OPE
\begin{equation}
  \dd\varphi(z)\dd\varphi(w)=\ffrac{1}{(z-w)^2}
\end{equation}
and the mode expansion
\begin{equation}\label{scalar-modes}
  \dd\varphi(z)=\sum_{n\in\oZ}\varphi_n z^{-n-1}.
\end{equation}
The energy--momentum tensor is given by
\begin{equation}\label{eq:the-Virasoro}
  T(z)=\half\,\dd\varphi(z)\dd\varphi(z)
  +\ffrac{\alpha_0}{2}\,\dd^2\varphi(z)
\end{equation}
(see~\eqref{eq:numbers}).  The modes of $\dd\varphi(z)$ span the
Heisenberg algebra and the modes of $T(z)$ span the Virasoro algebra
$\Vir$ with the central charge 
\begin{equation}\label{eq:centr-charge}
  c=1-6\ffrac{\bigl(p_+-p_-\bigr)^2}{p_+p_-}.
\end{equation}

The vertex operators are given by $e^{j(r,s)\varphi(z)}$ with $j(r,s)=
\frac{1-s}{2}\,\alpha_- + \frac{1-r}{2}\,\alpha_+$, \ $r,s\in\oZ$.
Equivalently, these vertex operators can be parameterized as
\begin{equation}\label{V-param}
  V_{r,s;n}(z)= e^{\frac{p_-(1-r)-p_+(1-s)+p_+p_-n}{\sqrt{2p_+p_-}}
    \varphi(z)},
  \qquad 1\leq r\leq p_+,\quad 1\leq s\leq p_-,\quad n\,{\in}\,\oZ.
\end{equation}
The conformal dimension of $V_{r,s;n}(z)$ assigned by the
energy--momentum tensor is
\begin{equation}\label{Delta-rs}
  \Delta_{r,s;n}=\ffrac{(p_+ s\!-\!p_- r\!+\!p_+ p_- n)^2
    -(p_+\!-\!p_-)^2}{4p_+p_-}.
\end{equation}
We note that
\begin{equation}\label{id-dim}
  \Delta_{-r,-s;-n}=\Delta_{r,s;n},\qquad
  \Delta_{r+kp_+,s+kp_-;n}=\Delta_{r,s;n},
  \qquad  \Delta_{r,s+kp_-;n}=\Delta_{r,s;n+k}.
\end{equation}

The vertex operators satisfy the braiding relations
\begin{multline}
  V_{r,s;n}(z_1)V_{r',s';n'}(z_2)=\\ 
  =\q^{(p_-(1-r)-p_+(1-s)+p_+p_-n)(p_-(1-r')-p_+(1-s')+p_+p_-n')}
   V_{r',s';n'}(z_2)V_{r,s;n}(z_1),
\end{multline}
where
\begin{equation}\label{small-q}
  \q=e^{\frac{i\pi}{2p_+p_-}}.
\end{equation}
(The convention is to take the OPE of the operators in the left-hand
side with $|z_1|>|z_2|$ and make an analytic continuation to
$|z_1|<|z_2|$ by moving $z_1$ along a contour passing below $z_2$,
i.e., as $z_1-z_2\to e^{i\pi}(z_1-z_2)$.)

In what follows, we also write $V_{r,s;0}(z)\equiv V_{r,s}(z)$ and
$\Delta_{r,s;0}\equiv\Delta_{r,s}$.

\subsection{Definition of modules}\label{subsec:modules}
For $1\leq r\leq p_+$, $1\leq s\leq p_-$, and $n\,{\in}\,\oZ$, let
$\repF_{r,s;n}$ denote the Fock module of the Heisenberg algebra
generated from (the state corresponding to) the vertex operator
$V_{r,s;n}(z)$.  The zero mode~$\varphi_0=\frac{1}{2i\pi}\oint
dz\dd\varphi(z)$ acts in~$\repF_{r,s;n}$ by multiplication with the
number
\begin{equation*}
  \varphi_0\,v=
  \ffrac{p_-(1-r)-p_+(1-s)+p_+p_-n}{
    \sqrt{\mathstrut 2p_+p_-}}\,v,
  \quad
  v\in\repF_{r,s;n}.
\end{equation*}
We write $\repF_{r,s}\equiv\repF_{r,s;0}$.  For convenience of
notation, we identify $\repF_{0,s;n}\equiv\repF_{p_+,s;n+1}$ and
$\repF_{r,0;n}\equiv\repF_{r,p_-;n-1}$.

Let $\rep{Y}_{r,s;n}$ with $1\leq r\leq p_+$, $1\leq s\leq p_-$, and
$n\,{\in}\,\oZ$ denote the Virasoro module that coincides with
$\repF_{r,s;n}$ as a linear space, with the Virasoro algebra action
given by~\eqref{eq:the-Virasoro} (see~\cite{[FF2]}).  As with the
$\repF_{r,s;n}$, we also write $\rep{Y}_{r,s}\equiv\rep{Y}_{r,s;0}$.

\subsubsection{Subquotient structure of the modules
  $\rep{Y}_{r,s;n}$}\label{subsec:FF-module} We recall the subquotient
structure of the Virasoro modules $\rep{Y}_{r,s;n}$~\cite{[FF]}.  We
let $\repJ_{r,s;n}$ denote the irreducible Virasoro module with the
highest weight~$\Delta_{r,s;n}$ (as before, $1\leq r\leq p_+$, $1\leq
s\leq p_-$, and $n\,{\in}\,\oZ$).  Evidently,
$\repJ_{r,s;n}\simeq\repJ_{p_+-r,p_--s;-n}$.  The
$\half(p_+\,{-}\,1)(p_-\,{-}\,1)$ nonisomorphic modules among the
$\repJ_{r,s;0}$ with $1\leq r\leq p_+\,{-}\,1$ and $1\leq s\leq
p_-\,{-}\,1$ are the irreducible modules from the Virasoro $(p_+,p_-)$
minimal model.  We also write $\repJ_{r,s}\equiv\repJ_{r,s;0}$. For
convenience of notation, we identify
$\repJ_{0,s;n}\equiv\repJ_{p_+,s;n+1}$ and
$\repJ_{r,0;n}\equiv\repJ_{r,p_-;n-1}$.

The well-known structure of $\rep{Y}_{r,s}$ for $1\leq r\leq
p_+\,{-}\,1$ and $1\leq s\leq p_-\,{-}\,1$ is recalled in
Fig.~\ref{fig:embedding}.
\begin{figure}[tbp]
  \mbox{}\hfill\mbox{}
  \xymatrix
  {{}&\times\ar[dr]\ar@{}|{[r,s;0]}[]+<20pt,20pt>&&\\
    \blacktriangle\ar[ur]\ar[d]\ar[drr]\ar@{}|{\kern-25pt
      \substack{[p_+-r,s;1]}}[]+<-30pt,10pt>
    &&\bullet\ar@{}|{\substack{[r,p_--s;1]\kern-10pt}}[]+<45pt,10pt>\\
    \circ\ar[urr]\ar[drr]\ar@{}|{[r,s;2]}[]+<-45pt,10pt>
    &&\scrBox\ar[u]\ar[d]\ar@{}|{[p_+-r,p_--s;2]}[]+<70pt,10pt>\\
    \blacktriangle\ar[u]\ar[d]\ar[urr]\ar[drr]
    \ar@{}|{\kern-7pt\substack{[p_+-r,s;3]}}[]+<-45pt,10pt>
    &&\bullet\ar@{}|{\substack{[r,p_--s;3]}\kern-7pt}[]+<45pt,10pt>\\
    \circ\ar[urr]\ar@{}|{[r,s;4]}[]+<-45pt,10pt> \ar@{.}[]-<0pt,18pt>
    &&\scrBox\ar[u]\ar@{}|{[p_+-r,p_--s;4]}[]+<70pt,10pt>
    \ar@{.}[]-<0pt,18pt>}
  \mbox{}\hfill\mbox{}
  \caption[Embedding structure of
  $\rep{Y}_{r,s}$]{\captionfont{Embedding structure of the
      Feigin--Fuchs module $\rep{Y}_{r,s}$}.  \footnotesize The
    notation is as follows.  The cross {\small$\times$} corresponds to
    the subquotient $\repJ_{r,s}$, the filled dots {\small$\bullet$}
    to $\repJ_{r,p_--s;2n+1}$ with $n\in\oN_0$, the triangles
    {\small$\blacktriangle$} to $\repJ_{p_+-r,s;2n+1}$
    with~$n\in\oN_0$, the open dots {\small$\circ$} to
    $\repJ_{r,s;2n}$ with~$n\in\oN$, and the squares {\small$\scrBox$}
    to $\repJ_{p_+-r,p_--s;2n}$ with $n\in\oN$. The arrows correspond
    to embeddings of the Virasoro modules and are directed toward
    submodules.  The notation $[a,b;n]$ in square brackets is for
    subquotients isomorphic to~$\repJ_{a,b;n}$.  The filled dots
    constitute the socle of $\rep{Y}_{r,s}$.}
  \label{fig:embedding}
\end{figure}

\subsubsection{}
The Fock spaces introduced above constitute a free-field module
\begin{equation}\label{lattice-F}
  \repF
  =\bigoplus_{n\in\oZ}
  \bigoplus_{r=1}^{p_+}\bigoplus_{s=1}^{p_-}\repF_{r,s;n}.
\end{equation}
It can be regarded as (the chiral sector of) the space of states of
the Gaussian Coulomb gas model compactified on the circle of
radius~$\sqrt{2p_+p_-}$.  This model has $c=1$ and its symmetry is a
lattice vertex operator algebra, which is described in the next
subsection.

\subsection{The lattice vertex-operator algebra}\label{sec:lvoa}
Let $\talgL$ be the lattice vertex-operator algebra
(see~\cite{[Kac],[FHL],[LL]}) generated by the vertex operators
\begin{equation*}
  V_{1,1;2n}(z) =
  e^{n\sqrt{2p_+p_-}\varphi(z)}, \quad n\,{\in}\,\oZ.
\end{equation*}
The underlying vector space (the vacuum representation) of $\talgL$ is
\begin{equation}\label{eq:vacLdec}
  \tvacL=\bigoplus_{n\in\oZ}\repF_{1,1;2n}.
\end{equation}
The vertex-operator algebra $\talgL$ has $2p_+p_-$ irreducible
modules, denoted in what follows as~$\repV_{r,s}^\pm$ with $1\leq
r\leq p_+$ and $1\leq s\leq p_-$. Their Fock-module decompositions are
given by
\begin{equation}\label{eq:w-verma-def}
  \begin{aligned}
    \repV_{r,s}^+&=\bigoplus_{n\in\oZ}\repF_{p_+-r,p_--s;2n},\\
    \repV_{r,s}^-&=\bigoplus_{n\in\oZ}\repF_{p_+-r,p_--s;2n+1},
  \end{aligned}
  \qquad 1\leq r\leq p_+,\quad 1\leq s\leq p_-.
\end{equation}
In this numerology, the vacuum representation $\tvacL$ coincides not
with the usual~$\repV_{1,1}^+$ but with~$\repV_{p_+-1,p_--1}^+$; such
a ``twist of notation'' turns out to be convenient in what follows.

The model with the space of states~\eqref{lattice-F} is rational with
respect to $\tvacL$, i.e., the space $\repF$ is a finite sum of
irreducible $\talgL$-modules:\pagebreak[3]
\begin{equation*}
  \repF =\bigoplus_{r=1}^{p_+}\bigoplus_{s=1}^{p_-}
  (\repV_{r,s}^+\oplus\repV_{r,s}^-).
\end{equation*}

\begin{figure*}[tbp]
  \begin{center}
    \includegraphics[bb=.7in 4.9in 7.8in 10.4in, clip,
    scale=0.8]{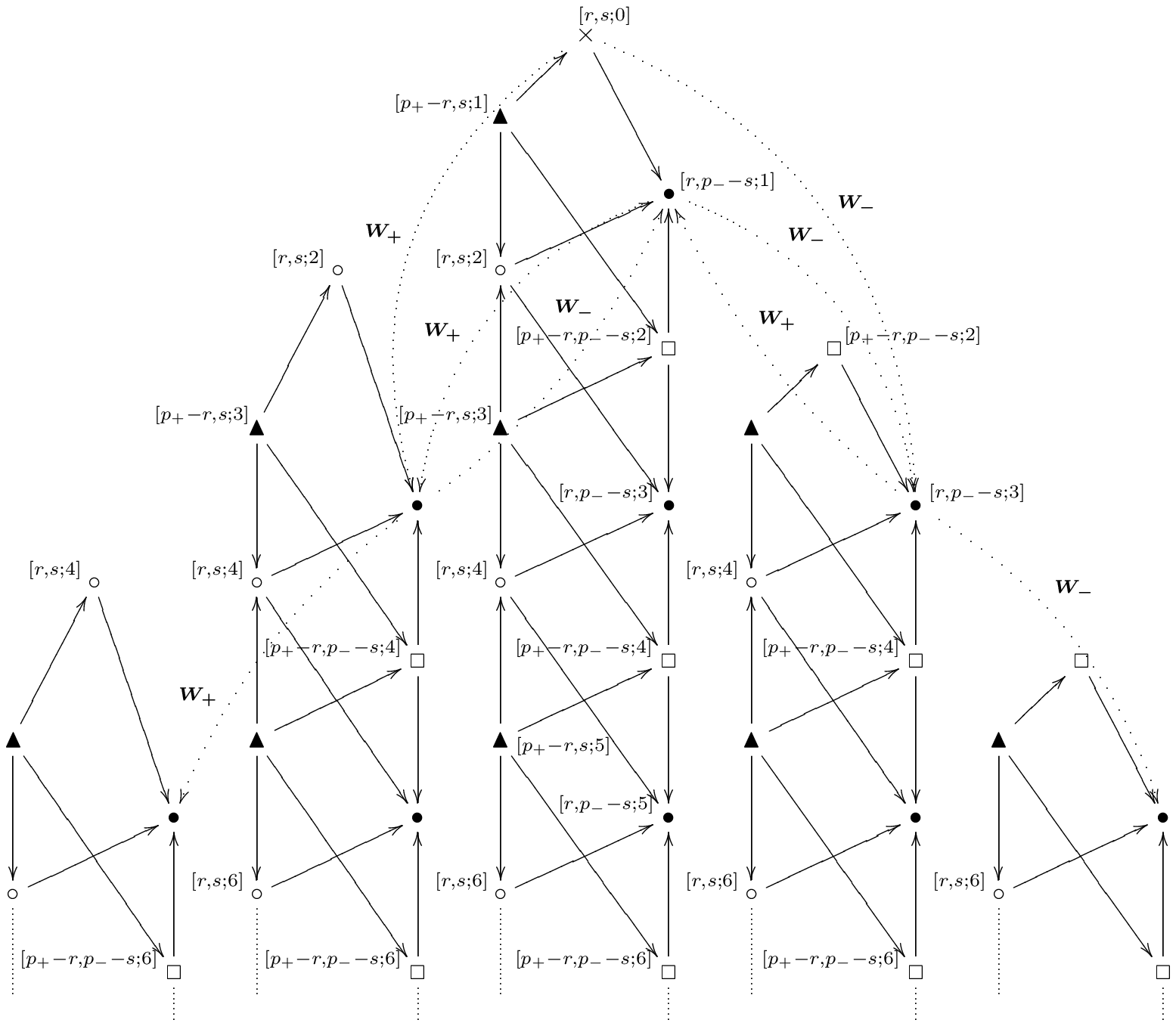}
   \footnotesize
    \mbox{}\hfill\mbox{}
    \caption[Structure of the modules
    $\repV_{p_+-r,p_--s}^+$]{\captionfont{Structure of the modules
        $\repV_{p_+-r,p_--s}^+$}.  \footnotesize The notation is the
      same as in Fig.~\ref{fig:embedding}.  Filled dots denote
      Virasoro representations that are combined
      into~$\repX_{r,s}^{+}$.  Levels of conformal dimensions are
      chosen for $(r,s)\in\setii$, and $p_->p_+$. The dotted lines
      show the action of the $W^\pm(z)$ generators of $\talgW$ in
      $\repK_{r,s}^+$, see Eq.~\eqref{eq:ker-scr} below.}
     \label{fig:W-action}
  \end{center}
\end{figure*}

\begin{figure}[tbhp]
  \begin{center}
    \includegraphics[scale=.8]{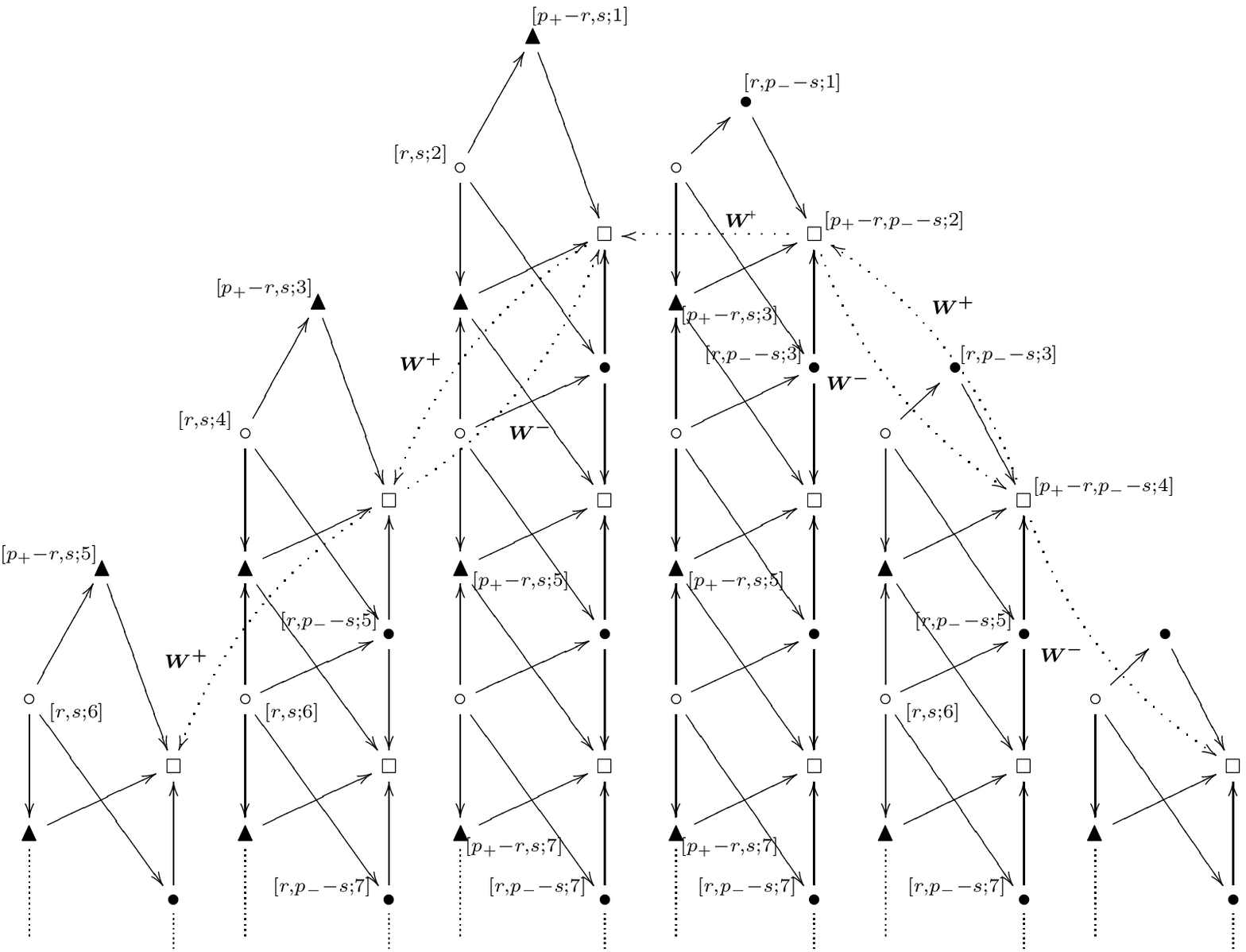}
    \caption[Structure of the modules
    $\repV_{r,p_--s}^-$]{\captionfont{Structure of the modules
        $\repV_{r,p_--s}^-$}.  \footnotesize The notation is the same
      as in Fig.~\ref{fig:embedding}. Boxes $\scrBox$ denote Virasoro
      representations that are combined into~$\repX_{p_+-r,s}^{-}$.
      Levels of conformal dimensions are chosen for $(r,s)\in\setii$,
      and $p_->p_+$. The dotted lines show the action of the
      $W^\pm(z)$ generators of $\talgW$ in $\repX_{p_+-r,s}^-$.}
   \label{fig:W-action-min}
  \end{center}
\end{figure}

Evidently, each $\repV_{r,s}^\pm$ is a Virasoro module by virtue of
the free-field construction~\eqref{eq:the-Virasoro}.  We need to
recall a number of standard facts known in the representation theory
of the Virasoro algebra.  With the same notation as in
Fig.~\ref{fig:embedding}, we describe the Virasoro structure of the
$\repV_{r,s}^\pm$ in Fig.~\ref{fig:W-action} ($\repV_{p_+-r,p_- -
  s}^+$), Fig.~\ref{fig:W-action-min} ($\repV_{r,p_--s}^-$),
Fig.~\ref{fig:W-action-r,p_--min} ($\repV_{r,p_-}^+$), and
Fig.~\ref{fig:W-action-r,p_-} ($\repV_{p_+-r,p_-}^-$).
\begin{figure}[tbhp]
  \begin{center}
    \includegraphics[bb=1.2in 6.9in 7in 10.4in, clip]{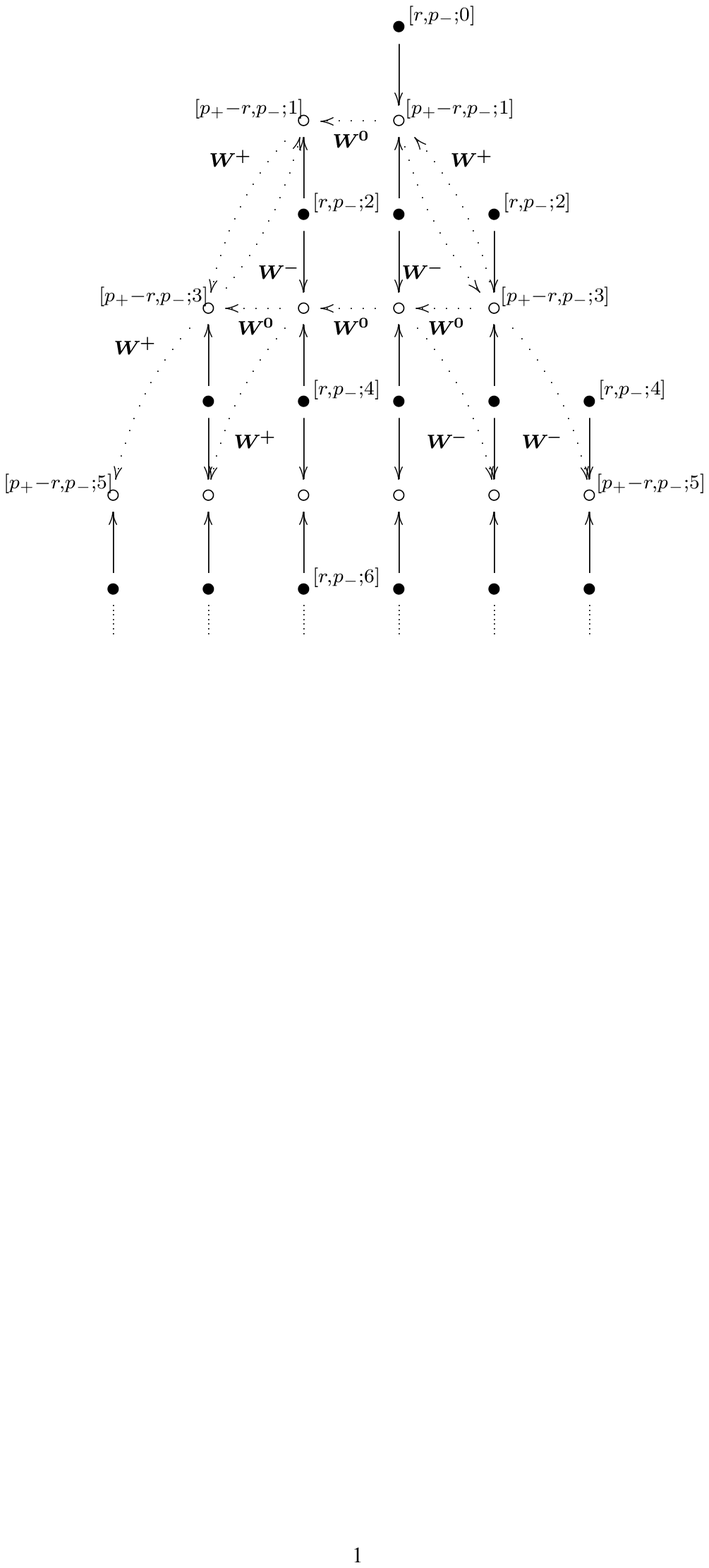}
    \caption[Structure of the modules
    $\repV_{r,p_-}^+$]{\captionfont{Structure of the modules
        $\repV_{r,p_-}^+$}.  \footnotesize The notation is the same as
      in Fig.~\ref{fig:embedding}.  Open dots~{\small $\circ$} denote
      Virasoro representations that are combined
      into~$\repX_{p_+-r,p_-}^{-}$.  The dotted lines show the action
      of the $W^\pm(z)$ generators of $\talgW$ in
      $\repX_{p_+-r,p_-}^-$.}
    \label{fig:W-action-r,p_--min}
  \end{center}
\end{figure}
\begin{figure}[tbhp]
  \begin{center}
    \includegraphics[bb=1.2in 6.9in 7in 10.4in, clip]{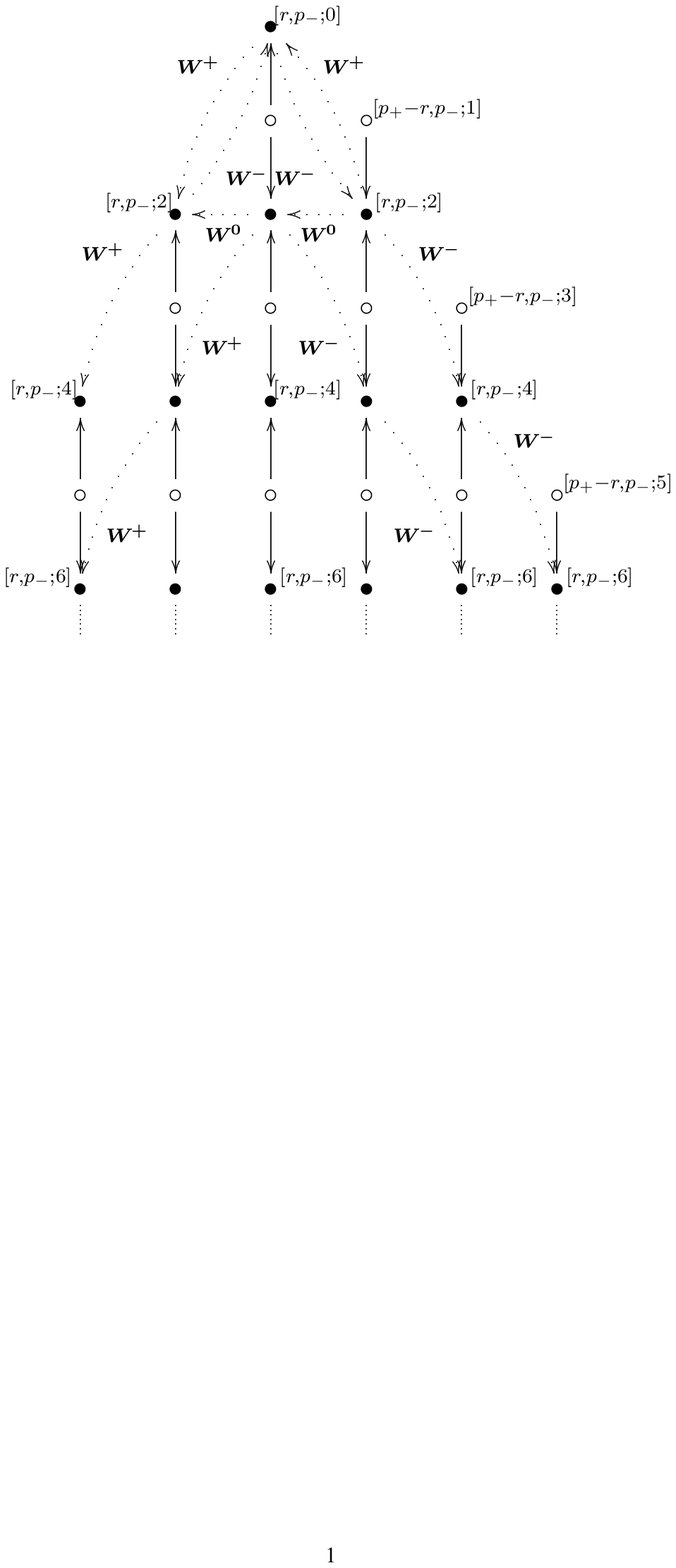}
    \caption[Structure of the modules
    $\repV_{p_+-r,p_-}^-$]{\captionfont{Structure of the modules
        $\repV_{p_+-r,p_-}^-$}.  \footnotesize The notation is the
      same as in Fig.~\ref{fig:embedding}. Filled dots~{\small
        $\bullet$} denote Virasoro representations that are combined
      into~$\repX_{r,p_-}^{+}$. The dotted lines show the action of
      the $W^\pm(z)$ generators of $\talgW$ in $\repX_{r,p_-}^+$.}
    \label{fig:W-action-r,p_-}
  \end{center}
\end{figure}

At the moment, the reader is asked to ignore the dotted lines in the
figures (they are to be described in Sec.~\ref{sec:w-rep}).  The
action of $\talgL$ in $\repV_{r,s}^\pm$ is as follows.  In
Figs.~\ref{fig:W-action} and~\ref{fig:W-action-min}, the Heisenberg
subalgebra acts in each two-strand ``braid'' (labeled by the last
integer in the square brackets at the top of each braid, even in
Fig.~\ref{fig:W-action} and odd in Fig.~\ref{fig:W-action-min}) as in
the corresponding Fock module.  Vertex operators $V_{1,1;2n}(z)$ map
between the braids over the distance~$n$ (to the left for $n>0$).  In
Figs.~\ref{fig:W-action-r,p_--min} and~\ref{fig:W-action-r,p_-}, the
Heisenberg subalgebra acts in each vertical strand and $V_{1,1;2n}(z)$
act between strands similarly.

We next consider the socle (the maximal semisimple submodule) of
$\repV_{r,s}^\pm$ (the irreducible~$\talgL$-module defined
in~\eqref{eq:w-verma-def}), viewed as a Virasoro module.

\begin{dfn}\label{dfn:repX}
  With the $\repV_{r,s}^{\pm}$ modules regarded as Virasoro modules,
  we set
  \begin{alignat*}{2}
    \repX^{\pm}_{r,s}&=\soc\repV^{\pm}_{p_+-r,p_--s},&\quad
    &1\leq r\leq p_+\!-\!1,\quad 1\leq s\leq p_-\!-\!1,\\
    \repX_{r,p_-}^\pm&=\soc\repV_{p_+-r,p_-}^\mp,&&1\leq r\leq
    p_+\!-\!1,\\
    \repX_{p_+,s}^\pm&=\soc\repV_{p_+,p_--s}^\mp,&&1\leq s\leq
    p_-\!-\!1,\\
    \repX_{p_+,p_-}^\pm&=\soc\repV_{p_+,p_-}^\pm.
  \end{alignat*}
\end{dfn}
We note that $\repX_{p_+,p_-}^\pm=\repV_{p_+,p_-}^\pm$.
Therefore, in particular,
\begin{equation}\label{socF}
  \soc\repF
  =\bigoplus_{r=1}^{p_+}\bigoplus_{s=1}^{p_-}
  \bigl(\repX_{r,s}^+\oplus\repX_{r,s}^-\bigr).
\end{equation}

The space $\repX_{r,s}^+$ is represented by the collection of filled
dots in Figs.~\ref{fig:W-action} and~\ref{fig:W-action-r,p_-},
$\repX_{p_+-r,s}^-$ by the collection of boxes in
Fig.~\ref{fig:W-action-min}, and $\repX^{-}_{p_+-r,p_-}$ by the
collection of open dots in Fig.~\ref{fig:W-action-r,p_--min}.  They
decompose into direct sums of irreducible Virasoro modules as follows.

\begin{lemma}\label{prop:x-vir}
  As a Virasoro module, the space $\repX^{\pm}_{r,s}$ for $1\leq r\leq
  p_+$ and $1\leq s\leq p_-$ decomposes as
  \begin{equation}\label{eq:w-vir-decomp}
    \repX^+_{r,s}\simeq\bigoplus_{a\geq0}(2a+1)
    \repJ_{r,p_--s;2a+1},\qquad
    \repX^-_{r,s}\simeq\bigoplus_{a\geq1}2a\repJ_{r,p_--s;2a},
  \end{equation}
  with the identification $\repJ_{0,s;n}\equiv\repJ_{p_+,s;n+1}$ and
  $\repJ_{r,0;n}\equiv\repJ_{r,p_-;n-1}$ introduced above.
\end{lemma}

As we see in what follows, the $\repX^\pm_{r,s}$ become $W$-algebra
representations.  Describing the $W$-algebra requires studying the
action of screenings, which we consider in the next section.

\section{Screening operators and the quantum
  group}\label{sec:screenings} In this section, we introduce screening
operators and study the quantum group associated with them.  Because
the screenings do not act in the free-field module~$\repF$, we have to
extend~$\repF$ to a larger space $\bigrepF$
in~\bref{subsec:screenings}.  In~\bref{subsec:q-gr}, we then
reformulate the action of the screenings as a representation of a Hopf
algebra, with the result formulated in~\bref{thm:double}.  A
subalgebra in a quotient of the Drinfeld double of this Hopf algebra
is the quantum group that is Kazhdan--Lusztig dual to the $W$-algebra.
In~\bref{sec:KerIm}, we next describe the spaces $\repX^\pm_{r,s}$
introduced above in terms of the screenings.  We also show
in~\bref{sec:Lusztig} that the relevant spaces carry a representation
of the~$s\ell(2,\oC)$ algebra commuting with the Virasoro action.  All
these ingredients are to be used in the next section in the study of
the $\WWW$ algebra.

\subsection{Screening operators and dressing}\label{subsec:screenings}
Free-field construction of both the minimal model and its logarithmic
extension involves screening operators $\ep$ and $\fm$ that commute
with the energy--momentum tensor, $[\ep,T(z)]=[\fm,T(z)]=0$.  They
have the standard form
\begin{equation}\label{screenings}
  \ep=\oint dz e^{\alpha_+\varphi(z)},\qquad
  \fm=\oint dz e^{\alpha_-\varphi(z)}.
\end{equation}

The operators $\ep$ and $\fm$ do not act in the space $\repF$
(see~\eqref{lattice-F}), and to complete their definition, we must
extend this space appropriately.  For this, we introduce the space
$\bigrepF$ spanned by dressed fields (cf.~\cite{[Bouw],[F]})
\begin{multline}\label{localsystem}
  \int_{\hat{\Gamma}} dz_1\dots dz_m dw_1\dots dw_j
  e^{\alpha_+\varphi(z_1)}\dots
  e^{\alpha_+\varphi(z_j)}e^{\alpha_-\varphi(w_1)}\dots\\
  {}\times e^{\alpha_-\varphi(w_m)} \mathcal{P}(\dd\varphi)
  V_{r,s;n}(z),\qquad
  0\leq j\leq p_+\!-\!1,\quad0\leq m\leq p_-\!-\!1,
\end{multline}
where $\hat{\Gamma}$ is a local system defined as in~\cite{[Bouw]} and
$\mathcal{P}(\dd\varphi)$ is a differential polynomial
in~$\dd\varphi$.

The action of $\ep$ and $\fm$ on each vector~\eqref{localsystem} can
then be evaluated by manipulations with contour integrals, as
described in~\cite{[Bouw],[F]}.  In particular,
\begin{alignat*}{2}
  \ep^r&:\repF_{r,s;n} &\to \repF_{p_+-r,s;n+1},\\
  \fm^s&:\repF_{r,s;n}&\to \repF_{r,p_--s;n-1}.
\end{alignat*}
Evidently, these spaces and the maps between them constitute the
Felder complexes~\cite{[F]} whose cohomology gives the standard
minimal model.

\subsection{The Hopf algebra of screenings}\label{subsec:q-gr}
The action of screening operators in~$\bigrepF$ gives rise to a Hopf
algebra representation.  The following lemma basically restates some
known facts about the screenings~\cite{[Bouw],[F]}.

\begin{lemma}
  The space $\bigrepF$ defined in~\bref{subsec:screenings} admits the
  action of the operators
  \begin{equation}
    \ep,\qquad
    \fm,\qquad
    \kk=e^{\frac{i\pi}{\sqrt{\mathstrut 2p_+p_-}}\varphi_0},
  \end{equation}
  which satisfy the relations
  \begin{equation}\label{B-relations}
    \begin{gathered}
      \ep^{p_+}=\fm^{p_-}=0,\quad\kk^{4p_+p_-}=\one,
      \quad\ep\fm=\fm\ep,\\
      \kk\ep\kk^{-1}=\qp\ep,\quad\kk\fm\kk^{-1}=\qm^{-1}\fm,
    \end{gathered}
  \end{equation}
  where \textup{(}see~\eqref{small-q}\textup{)}
  \begin{equation*}
    \qp=\q^{2p_-}=e^{\frac{i\pi}{p_+}},\quad
    \qm=\q^{2p_+}=e^{\frac{i\pi}{p_-}}.
  \end{equation*}
\end{lemma}

Let $\qalgB$ denote the associative algebra generated by $\ep$, $\fm$,
and $\kk$ with relations~\eqref{B-relations}.  Clearly, the PBW basis
in $\qalgB$ is given by
\begin{equation}\label{ejmn}
  \pbw_{jmn}=\kk^j\ep^m\fm^n,
  \qquad 0\leq j\leq 4p_+p_-\,{-}\,1,
  \quad 0\leq m\leq p_+ -1,
  \quad 0\leq n\leq p_- -1.
\end{equation}
\begin{lemma}
  The algebra $\qalgB$ is a Hopf algebra with the counit
  \begin{equation}
    \epsilon(\ep)=\epsilon(\fm)=0,\quad\epsilon(\kk)=1,
  \end{equation}
  comultiplication
  \begin{equation}
    \Delta(\kk)=\kk\tensor\kk,
    \quad
    \Delta(\ep) =\ep\tensor\one+\kk^{2p_-}\tensor\ep,
    \quad
    \Delta(\fm)=\fm\tensor\one+\kk^{-2p_+}\tensor\fm,
  \end{equation}
  and antipode
  \begin{equation}\label{the-antipode}
    S(\kk)=\kk^{-1},\quad S(\ep)=-\kk^{-2p_-}\ep,\quad
    S(\fm)=-\kk^{2p_+}\fm.
  \end{equation}
\end{lemma}

We note that the counit is given by the trivial representation on the
vertex $V_{1,1}=\one$.  The comultiplication is taken from operator
product expansion; for example, the standard manipulations lead to
\begin{multline*}
  \ep\Bigl(e^{j_1\varphi(w)}e^{j_2\varphi(u)}\Bigr)= \oint
  e^{\alpha_+\varphi(z)}
  \Bigl(e^{j_1\varphi(w)}e^{j_2\varphi(u)}\Bigr)dz=\\
  {}=\Bigl(\oint e^{\alpha_+\varphi(z)}
  e^{j_1\varphi(w)}dz\Bigr)e^{j_2\varphi(u)}
  +e^{i\pi\alpha_+j_1}e^{j_1\varphi(w)}
  \Bigl( \oint e^{\alpha_+\varphi(z)}e^{j_2\varphi(u)}dz\Bigr)=\\
  {}=\ep
  e^{j_1\varphi(w)}e^{j_2\varphi(u)}+\kk^{2p_-}e^{j_1\varphi(w)}\ep
  e^{j_2\varphi(u)},
\end{multline*}
which gives the comultiplication for $\ep$, and similarly for $\fm$
and $\kk$.  With the counit and comultiplication thus fixed, the
antipode $S$ is uniquely calculated from the comultiplication, with
the result in~\eqref{the-antipode}.

The full quantum group realized in a given conformal field theory
model involves not only the screening but also the contour-removal
operators.  A convenient procedure for introducing the latter is to
take Drinfeld's double (see~\cite{[Kassel],[ChP]}
and~\bref{app:double}) of the algebra $\qalgB$ generated by the
screenings.

\begin{Thm}\label{thm:double}
   \addcontentsline{toc}{subsection}{\thesubsection.  \ \
   Structure of the double}
  The double $D(\qalgB)$ of $\qalgB$ is a Hopf algebra generated by
  $\epm$, $\fpm$, $\kk$, and $\dkk$ with the relations
  \begin{gather}
    \epm^{p_\pm}=\fpm^{p_\pm}=0,\quad\kk^{4p_+p_-}=\dkk^{4p_+p_-}=\one,\\
    \kk\epm\kk^{-1}=\qpm\epm,\quad\kk\fpm\kk^{-1}=\qpm^{-1}\fpm,\quad
    \dkk\epm\dkk^{-1}=\qpm^{-1}\epm,\quad\dkk\fpm\dkk^{-1}=\qpm\fpm,\\
    \kk\dkk=\dkk\kk,\quad\ep\emi=\emi\ep,\quad\fp\fm=\fm\fp,
    \quad\ep\fm=\fm\ep,\quad
    \emi\fp=\fp\emi,\\
    {[}\epm,\fpm] =\ffrac{\kk^{\pm2p_\mp}-\dkk^{\pm2p_\mp}}{
      \qpm^{\pm p_\mp}\!- \qpm^{\mp p_\mp}},\\
    \Delta(\kk)\!=\!\kk\!\tensor\!\kk, \quad
    \Delta(\ep)\!=\!\ep\!\tensor\!\one+\kk^{2p_-}\!\tensor\!\ep, \quad
    \Delta(\fm)\!=\!\fm\!\tensor\!\one+\kk^{-2p_+}\!\tensor\!\fm,
    \label{Delta-double}\\
    \Delta(\dkk)\!=\!\dkk\!\tensor\!\dkk, \quad
    \Delta(\fp)\!=\!\fp\!\tensor\!\dkk^{2p_-}+\one\!\tensor\!\fp,
    \quad
    \Delta(\emi)\!=\!\emi\!\tensor\!\dkk^{-2p_+}+\one\!\tensor\!\emi,
    \\
    S(\kk)=\kk^{-1},\quad S(\ep)
    =-\kk^{-2p_-}\ep,\quad S(\fm)=-\kk^{2p_+}\fm,\\
    S(\dkk)=\dkk^{-1},\quad S(\fp)=-\fp\dkk^{-2p_-},
    \quad S(\emi)=-\emi\dkk^{2p_+},\\
    \epsilon(\epm)=\epsilon(\fpm)=0,\quad\epsilon(\kk)=\epsilon(\dkk)=1.
    \label{epsilon-double}
  \end{gather}
\end{Thm}
We prove this in~\bref{app:double-proof} by a routine application of
the standard construction.

The double thus introduces contour-removal operators $\emi$ and $\fp$,
dual to $\ep$ and $\fm$ in the sense that is fully clarified in the
proof in Appendix~\ref{app:QG}.  But the doubling procedure also
yields the dual $\dkk$ to the Cartan element $\kk$ in $\qalgB$,
which is to be eliminated by passing to the quotient
\begin{equation}\label{D-bar}
  \bar D(\qalgB)=D(\qalgB)\!\bigm/\!(\kk\dkk-\one)
\end{equation}
over the \textit{Hopf} ideal generated by the central element
$\kk\dkk-\one$.  We next take a subalgebra in $\bar D(\qalgB)$ (which,
unlike $\bar D(\qalgB)$, is a factorizable ribbon quantum group,
see~\cite{[FGST-q]}).
\begin{dfn}
  Let $\tqalgA$ be the subalgebra in $\bar D(\qalgB)$ generated by
  $\ep$, $\fp$, $\emi$, $\fm$, and
  \begin{equation}\label{K}
    K=k^2.
  \end{equation}
\end{dfn}
Explicit relations among the generators are a mere rewriting of the
corresponding formulas in the theorem,
\begin{gather*}
  \epm^{p_\pm}=\fpm^{p_\pm}=0,\quad K^{2p_+p_-}=\one,\\
  K\epm K^{-1}=\qpm^2\epm,\quad K\fpm K^{-1}=\qpm^{-2}\fpm,\\
  \ep\emi=\emi\ep,\quad\fp\fm=\fm\fp, \quad\ep\fm=\fm\ep,\quad
  \emi\fp=\fp\emi,\\
  {[}\epm,\fpm] =\ffrac{K^{\pm p_\mp} - K^{\mp p_\mp}}{ \qpm^{\pm
      p_\mp}\!- \qpm^{\mp p_\mp}},
\end{gather*}
with the comultiplication, antipode, and counit defined
in~\eqref{Delta-double}--\eqref{epsilon-double}.  The structure
of~$\tqalgA$ and its relation to the logarithmic $(p_+,p_-)$ model are
investigated in~\cite{[FGST-q]}.

This quantum group has $2p_+ p_-$ irreducible representations, which
we label as $\XX^{\pm}_{r,s}$ with $1\leq r\leq p_+$ and $1\leq s\leq
p_-$.  The $\tqalgA$-module $\XX^{\pm}_{r,s}$ is generated by~$\ep$
and~$\fm$ from the eigenvector of~$K$ with the eigenvalue
$\pm\qp^{r-1}\qm^{-s+1}$, and we have $\dim\XX^{\pm}_{r,s}=r s$.  In
what follows, we use the $\XXX$-modules to describe the bimodule
structure of~$\bigrepF$.

\subsection{Kernels and images of the screenings}\label{sec:KerIm} Our
aim in what follows is to introduce some other module structures in
the spaces $\repX^\pm_{r,s}$, so far defined as Virasoro modules
in~\bref{dfn:repX}.  For this, we first describe them an intersection
of the images of screenings.

The socle of the space $\repF$ (see~\eqref{socF}), still viewed as a
Virasoro module, can also be written as
\begin{align*}
  \soc\repF
  =\im\ep^{p_+-1}\cap\im\fm^{p_--1}\subset\bigrepF
\end{align*}
Equivalently (and somewhat more convenient technically), the spaces
$\repX_{r,s}^\pm$ that constitute $\soc\repF$ are described as
intersections of the images of the screenings as
\begin{equation}\label{eq:soc-V}
  \begin{alignedat}{4}
    \repX_{r,s}^\pm&=\im\ep^{p_+-r}\cap\im\fm^{p_--s}  
    &\quad &\text{in}\ \ \repV_{p_+-r,p_--s}^\pm\ \ &&\text{for}\
    &&\kern-4pt\begin{array}[t]{l}
      1\leq r\leq p_+\!-\!1,\\
      1\leq s\leq p_-\!-\!1,
    \end{array}\\
    \repX_{r,p_-}^\pm&=\im\ep^{p_+-r}
    &&\text{in}\ \ \repV_{p_+-r,p_-}^\mp\ &&\text{for}\ \
    &&1\leq r\leq p_+\!-\!1,\\
    \repX_{p_+,s}^\pm&=\im\fm^{p_--s}
    &&\text{in}\ \ \repV_{p_+,p_--s}^\mp\ &&\text{for}\ \
    &&1\leq s\leq p_-\!-\!1.
  \end{alignedat}
\end{equation}
In Fig.~\ref{fig:W-action}, the image of $\ep^{p_+-r}$ is given by
with the collection of filled and open dots and the image of
$\fm^{p_--s}$ by the collection of filled dots and boxes.

In addition to the images, we also consider the kernels, namely,
$\ker\ep\cap\ker\fm\subset\bigrepF$, which decomposes similarly
to~\eqref{socF} as
\begin{equation*}
  \ker\ep\cap\ker\fm=\bigoplus_{r=1}^{p_+}\bigoplus_{s=1}^{p_-}
  \bigl(\repK_{r,s}^+\oplus\repK_{r,s}^-\bigr),
\end{equation*}
where the $\repK_{r,s}^\pm$ can equivalently be identified in
$\repV_{p_+-r,p_--s}^\pm$ as
\begin{equation}\label{eq:ker-scr}
  \begin{alignedat}{4}
    \repK_{r,s}^\pm&=\ker\ep^{r}\cap\ker\fm^{s}
    &\quad &\text{in}\ \ \repV_{p_+-r,p_--s}^\pm\ \ &&\text{for}\ 
    &&\kern-5pt\begin{array}[t]{l}
      1\leq r\leq p_+\!-\!1,\\
      1\leq s\leq p_-\!-\!1,
    \end{array}\\
    \repK_{r,p_-}^\pm&=\ker\ep^{r}
    &&\text{in}\ \ \repV_{p_+-r,p_-}^\mp\ &&\text{for}\ \
    &&1\leq r\leq p_+\!-\!1,\\
    \repK_{p_+,s}^\pm&=\ker\fm^{s}
    &&\text{in}\ \ \repV_{p_+,p_--s}^\mp\ &&\text{for}\ \
    &&1\leq s\leq p_-\!-\!1
  \end{alignedat}
\end{equation}
(and $\repK_{p_+,p_-}^\pm=\repV_{p_+,p_-}^\pm$ for uniformity of
notation).  Clearly, the $\repK_{r,s}^\pm$ are Virasoro modules.  

In Fig.~\ref{fig:W-action}, the kernel of $\ep^{r}$ coincides with the
collection of filled and open dots and the cross and the kernel of
$\fm^{s}$ coincides with the collection of filled dots, boxes, and the
cross.  It is easy to see that
\begin{align}
  \label{K-reducile}
  &\repK_{r,s}^+\supset\repX_{r,s}^+
  \ \ \text{with}\ \ \repK_{r,s}^+/\repX_{r,s}^+=\repJ_{r,s}
  \ \ \text{for}\ \ 1\leq r\leq p_+\!-\!1,\quad
  1\leq s\leq p_-\!-\!1,\\
  &\repK_{r,s}^-=\repX_{r,s}^-\quad\text{for}\quad 
  1\leq r\leq p_+,\quad 1\leq s\leq p_-,\\
  &\repK_{r,s}^+=\repX_{r,s}^+\quad\text{whenever}
  \quad r=p_+\quad\text{or}\quad s=p_-.
\end{align}

In what follows, we see that the spaces $\repX^\pm_{r,s}$ are in fact
irreducible $\WWW$-represen\-ta\-tions; on the other hand, the
$\repK^\pm_{r,s}$, which are also $\WWW$-representations, are not all
irreducible.  However, they play a crucial role in the $\WWW$
representation theory; $\repK^+_{1,1}$ is the vacuum representation of
the $\WWW$ algebra, and the $\repK^\pm_{r,s}$ can be identified with
the preferred basis of the $\WWW$ fusion algebra~$\Grring$.

\subsection{A Lusztig extension of the quantum
  group}\label{sec:Lusztig}
Virasoro modules $\repX^\pm_{r,s}$ and $\repK^\pm_{r,s}$ admit an
action of the~$s\ell(2,\oC)$
algebra.
\begin{lemma}\label{lemma:sl2c}
  The spaces $\repX_{r,s}^\pm$ and $\repK_{r,s}^\pm$ admit an
  $s\ell(2,\oC)$-action that is a derivation of the operator product
  expansion.  The Virasoro algebra $\smash[b]{\Vir}$,
  see~\eqref{eq:the-Virasoro}, commutes with this~$s\ell(2,\oC)$.
\end{lemma}

The construction of the $s\ell(2,\oC)$ action is based on Lusztig's
divided powers and can be briefly described as
follows~\cite{[FeiginTipunin]}.  Morally, the $e$ and $f$ generators
of $s\ell(2,\oC)$ are given by $\ep^{p_+}/[p_+]_+!$ and
$\fm^{p_-}/[p_-]_-!$, where we use the standard notation
\begin{equation*}
  \qint{n}_\q = \ffrac{\q^{n}-\q^{-n}}{\q-\q^{-1}},\quad
  \qint{n}_\q! = \qint{1}_\q\qint{2}_\q\dots
  \qint{n}_\q
\end{equation*}
for $\q$-integers and $\q$-factorials and set
$\qint{m}_+=\qint{m}_{\qp^{p_-}}$ and
$\qint{m}_-=\qint{m}_{\qm^{p_+}}$.  But if taken literally, both
$\ep^{p_+}/[p_+]_+!$ and $\fm^{p_-}/[p_-]_-!$ are given by $0/0$
ambiguities for $\ep$ and $\fm$ acting as described above and for $\q$
in Eq.~\eqref{small-q} (because $[p_\pm]_\pm=0$).  To resolve the
ambiguities, we consider a deformation of $\q$ in Eq.~\eqref{small-q}
as
\begin{equation*}
  \q_\epsilon=e^{\frac{i\pi}{2p_+p_-}+\epsilon}
\end{equation*}
or, equivalently, a deformation of the parameter~$\alpha_0$.  We thus
obtain $\ep(\epsilon)$, $\fm(\epsilon)$, and $q$-factorials
$[n]^\epsilon_\pm!$ depending on $\epsilon$, such that
$[p_\pm]^\epsilon_\pm!\neq0$.  The limits
\begin{equation*}
  e=\lim_{\epsilon\to0}\ffrac{\ep(\epsilon)^{p_+}}{[p_+]^\epsilon_+!}
  \quad
  \text{and}\quad
  f=\lim_{\epsilon\to0}\ffrac{\fm(\epsilon)^{p_-}}{[p_-]^\epsilon_-!}
\end{equation*}
act in $\repK_{r,s}^\pm$ and are independent of the deformation
details, and are therefore well-defined operators in
$\repK_{r,s}^\pm$.  Because $e$ and $f$ commute with the Virasoro
algebra action in~$\repK_{r,s}^\pm$, the spaces $\repX_{r,s}^\pm$ are
also representations of the~$s\ell(2,\oC)$ algebra generated by $e$
and $f$.

We also note that $\repJ_{r,s}=\repK_{r,s}^+/\repX_{r,s}^+$ (where
$1\leq r\leq p_+\,{-}\,1$ and $1\leq s\leq p_-\,{-}\,1$) are
$s\ell(2,\oC)$-singlets.

\begin{lemma}\label{lemma:sl-vir}
  The spaces $\repK_{r,s}^\pm$ and $\repX_{r,s}^\pm$ have the
  structure of $(\smash[b]{\Vir},\!s\ell(2,\!\oC))$-bimod\-ules given
  by
  \begin{equation}\label{eq:vacuum-space}
    \repK^+_{r,s}=\smash[b]{\bigoplus_{n\in\oN}}\,\rep{C}_n\tensor\ell_{2n-1},
   \quad 1\leq r\leq p_+-1,\quad1\leq s\leq p_--1
  \end{equation}
  and
  \begin{equation}\label{eq:vir-sl}
    \begin{aligned}
      \repX^+_{r,s}&\simeq
      \bigoplus_{n\in\oN}\repJ_{r,p_--s;2n-1}\tensor\ell_{2n-1},\\
      \repX^-_{r,s}&\simeq
      \bigoplus_{n\in\oN}
      \repJ_{r,p_--s;2n}\tensor\ell_{2n},
    \end{aligned}
    \quad 1\leq r\leq p_+,\quad1\leq s\leq p_-,
  \end{equation}  
  where $\ell_n$ is the $n$-dimensional irreducible $s\ell(2,\oC)$
  representation and
  \begin{equation*}
    \rep{C}_n=
    \begin{cases}
      \repJ_{r,p_- - s;2n-1},& n\geq2,\\[2pt]
      \overset{{\mbox{\small$\repJ_{r,s}$}}}{\times}\longrightarrow
      \overset{\mbox{\small$\repJ_{r,p_--s;1}$}}{\bullet},&
      n=1,
    \end{cases}
  \end{equation*}
  where we use the same notation as in Fig.~\ref{fig:embedding} for
  the extension of the Virasoro module $\repJ_{r,s}$ by the Virasoro
  module~$\repJ_{r,p_--s;1}$.
\end{lemma}

Each direct summand in the decomposition of $\repX^+_{r,s}$
in~\eqref{eq:vir-sl} corresponds to a horizontal row of filled dots in
Figs.~\ref{fig:W-action} and~\ref{fig:W-action-r,p_-}, and each direct
summand in the decomposition of $\repX^-_{p_+-r,s}$ corresponds to a
horizontal row of boxes and open dots in Figs.~\ref{fig:W-action-min}
and~\ref{fig:W-action-r,p_--min} respectively, with the $s\ell(2,\oC)$
algebra acting in each row.  We thus have $1$, $3$, $5$,
\dots-dimensional $s\ell(2,\oC)$-representations in
Figs.~\ref{fig:W-action} and~\ref{fig:W-action-r,p_-}, and $2$, $4$,
$6$, \dots-dimensional representations in Figs.~\ref{fig:W-action-min}
and~\ref{fig:W-action-r,p_--min}.

\medskip

We now have all the ingredients needed for constructing the
$W$-algebra~$\talgW$ of the $(p_+,p_-)$ logarithmic conformal field
theory model.

\section{Vertex-operator algebra for the $(p_+,p_-)$-conformal field
  theory and its representations}\label{sec:w-rep} In this section, we
define the chiral symmetry algebra $\talgW$ of the $(p_+,p_-)$
logarithmic model and describe irreducible and Verma $\talgW$-modules.

\begin{Dfn}\label{subsec:w-def}
  \addcontentsline{toc}{subsection}{\thesubsection.  \ \ Definition of
    $\smash{\talgW}$}The algebra $\talgW$ is the subalgebra of
  $\talgL$ with the underlying vector space
  $\tvacW=\repK_{1,1}^+\subset\tvacL$ (see~\eqref{eq:ker-scr}).
\end{Dfn}

The $(\Vir, s\ell(2,\oC))$-bimodule structure of the vacuum
$\talgW\!$-representation $\tvacW$ is described in~\bref{lemma:sl-vir}
and can be understood better using Fig.~\ref{fig:W-action}, as the
part of the figure consisting of the cross and filled
dots.\footnote{In the Coulomb-gas picture, $\talgW$ can be viewed as
  the symmetry algebra of a fixed-point system with quenched
  disorder.}

\begin{Thm}\mbox{}\label{prop:w-alg}
  \addcontentsline{toc}{subsection}{\thesubsection.  \ \ The $\algW$
    algebra of $(p_+,p_-)$ logarithmic models.}
  \begin{enumerate}
  \item The algebra $\talgW$ is generated by $T(z)$
    in~\eqref{eq:the-Virasoro} and the two currents $W^+(z)$ and
    $W^-(z)$ given by
    \begin{equation}\label{W-pm}
      W^+(z)=(\fm)^{p_--1}V_{1,p_--1;3}(z),\qquad
      W^-(z)=(\ep)^{p_+-1}V_{p_+-1,1;-3}(z),
    \end{equation}
    which are Virasoro primaries of conformal dimension
    $(2p_+{-}1)(2p_-\,{-}\,1)$.
    
  \item The maximal \textup{(}and the only nontrivial\textup{)}
    vertex-operator ideal $\algR$ of $\talgW$ is generated by $W^+(z)$
    and $W^-(z)$.
    
  \item The quotient $\talgW/\algR$ is the vertex-operator
    algebra~$\talgM$ of the $(p_+,p_-)$ Virasoro minimal model.
    
  \end{enumerate}
\end{Thm}
\begin{rem}
  In accordance with the definition of the screening action
  in~\eqref{W-pm}, the $W^\pm(z)$ currents can be written as (with
  normal ordering understood)
  \begin{align*}
    W^+(z)&=\polP_{3p_+p_--3p_+-p_-+1}^+\,e^{p_+\alpha_+\varphi}(z),\\
    W^-(z)&=\polP_{3p_+p_--p_+-3p_-+1}^-\,e^{p_-\alpha_-\varphi}(z),
  \end{align*}
  where $\polP_j^\pm$ are polynomials of degree $j$ in $\dd^n\varphi$,
  $n\geq1$ (with~$\mathrm{deg}\,\dd^n\varphi=n$).  The OPE of $W^+(z)$
  and $W^-(z)$ is
  \begin{equation}
    W^+(z)W^-(w)=\frac{S_{p_+,p_-}(T)}{(z-w)^{7p_+p_--3p_+-3p_-+1}}+
    \text{less singular terms},
  \end{equation}
  where $S_{p_+,p_-}(T)$ is the vacuum singular vector\,---\,the
  polynomial of degree $\half
  (p_+\,{-}\,1)\cdot$\linebreak[0]$(p_-\,{-}\,1)$ in $T$ and $\dd^nT$,
  $n\geq1$, such that $S_{p_+,p_-}(T)=0$ is the polynomial relation
  for the energy--momentum tensor in the $(p_+,p_-)$ Virasoro minimal
  model.  We note that this OPE is already quite difficult for direct
  analysis in the simplest case of~$(3,2)$ model, see
  Appendix~\ref{app:32-example}.
\end{rem}

\begin{rem}
  As follows from~\bref{subsec:w-def} and~\bref{lemma:sl-vir}, $\talgW$
  admits an $s\ell(2,\oC)$ action and the $W^\pm(z)$ generators are
  highest- and lowest-weight vectors of an $s\ell(2,\oC)$-triplet.  We
  note that the $W$ algebra of $(p,1)$ logarithmic models bears the
  name \textit{triplet}~\cite{[Kausch-sympl],[GaberdielKausch1]}
  because it admits such an $s\ell(2,\oC)$ action and the
  corresponding $W^\pm(z)$ generators are also the highest- and
  lowest-weight vectors in a triplet.
\end{rem}

\begin{proof}[Proof of~\bref{prop:w-alg}]
  With decomposition \eqref{eq:vacuum-space} taken into account, it
  suffices to construct
  $(\Vir,$\linebreak[0]$s\ell(2,\oC))$ highest-weight vectors in each
  direct summand in order to show part~(1).  This can be done as
  follows.  The highest-weight vector of each component
  $C_n\tensor\ell_{2n-1}$ (corresponding to the rightmost dot in each
  row of filled dots in Fig.~\ref{fig:W-action}) is identified with
  the field $W^{-,n}(z)$, $n\geq1$, defined as the first nonzero term
  in the OPE of $W^{-}(z)$ with $W^{-,n-1}(z)$; the recursion base
  is~$W^{-,1}\equiv\one$.  The highest-weight vector of
  $C_1\tensor\ell_{1}$ is identified with the identity operator
  $\one$.  This shows part~(1).
  
  Next, it follows from~\bref{lemma:sl-vir} that
  $\repX_{1,1}^+=\im\ep^{p_+ - 1}\cap\im\fm^{p_- - 1}$ is invariant
  under $\talgW$.  The space $\repX_{1,1}^+$ is the maximal
  $\talgW$-submodule in $\tvacW=\repK_{1,1}^+$, and therefore there
  exists a map $\talgW\to\talgM$ with the kernel~$\repX_{1,1}^+$.  We
  set $\algR=\repX_{1,1}^+$, which is the maximal vertex-operator
  ideal in $\talgW$.  Evidently, $W^\pm(z)$ generate some subspace
  $\algR'$ in $\algR$, but the decomposition into the direct sum of
  $s\ell(2,\oC)$-representations allows calculating the character of
  $\algR'$, which coincides with the character of $\im
  \ep^{p_+\,{-}\,1}\cap\im \fm^{p_-\,{-}\,1}$ calculated from the
  Felder resolution, thus showing parts~(2) and~(3).
\end{proof}

\subsection{Irreducible $\smash{\talgW}$-modules}\label{subsec:w-mod}
We now describe irreducible $\talgW$-modules.  First, the irreducible
\textit{Virasoro} modules $\repJ_{r,s}$ with $(r,s)\in\setii$ are at
the same time $\talgW$-modules, with the ideal $\algR$ acting by zero.
\textit{We write $\repX_{r,s}$ for the $\repJ_{r,s}$ regarded as
  $\talgW$-modules}.  Second, there are $2p_+ p_-$ irreducible
$\talgW$-modules where $\algR$ acts nontrivially.  These are the
spaces $\repX^{\pm}_{r,s}$ introduced in~\bref{dfn:repX}, which are
evidently $\talgW$-modules because of their description
in~\eqref{eq:soc-V}.

\begin{prop}
  $\repX^{\pm}_{r,s}$ is an irreducible $\talgW$-module.
\end{prop}
\begin{proof}
  Taking decomposition~\eqref{eq:vir-sl} into account, we show the
  irreducibility of $\repX^{\pm}_{r,s}$ by literally repeating the
  construction for the $(\smash[b]{\Vir},s\ell(2,\oC))$ highest-weight
  vectors in the proof of~\bref{prop:w-alg}.
\end{proof}

The lowest-dimension Virasoro primary in~$\repX^+_{r,s}$
is~$V_{r,p_--s;1}$ (of dimension $\Delta_{r,p_--s;1}$) and two
lowest-dimension Virasoro primaries in~$\repX^-_{r,s}$ are
$(\ep)^{p_+-r}V_{p_+-r,s;-2}$ and
$(\fm)^{p_--s}$\linebreak[0]$V_{r,p_--s;2}$ (of dimension
$\Delta_{p_+-r,s;-2}$).  The irreducible representations can be
arranged into a $p_+\times p_-$ Kac table, with $\repX^{+}_{r,s}$ and
$\repX^{-}_{r,s}$ in each box (in addition, the standard
$(p_+\,{-}\,1)\times(p_-\,{-}\,1)$ Kac table containing
$\half(p_+\,{-}\,1)(p_-\,{-}\,1)$ distinct representations is
inherited from the minimal model); the $(3,2)$ example is given in
Table~\ref{table-kac}.  The $\talgW$-action in $\repX^{\pm}_{r,s}$ is
shown by dotted lines in
Figs.~\ref{fig:W-action}--\ref{fig:W-action-r,p_-}: for
$\repX^{+}_{r,s}$ with $1\leq r\leq p_+\,{-}\,1$ and $1\leq s\leq
p_-\,{-}\,1$ in Fig.~\ref{fig:W-action}, for $\repX^{-}_{p_+-r,s}$
with $1\leq r\leq p_+\,{-}\,1$ and $1\leq s\leq p_-\,{-}\,1$ in
Fig.~\ref{fig:W-action-min}, for $\repX^{-}_{p_+-r,p_-}$ with $1\leq
r\leq p_+\,{-}\,1$ in Fig.~\ref{fig:W-action-r,p_--min}, and for
$\repX^{+}_{r,p_-}$ with $1\leq r\leq p_+\,{-}\,1$ in
Fig.~\ref{fig:W-action-r,p_-}.
\begin{table}[tbph]
  \centering
  \caption[Kac table]{\captionfont{The $p_+\times p_-$ $W$-algebra Kac
      table for $p_+=3$ and $p_-=2$}. \footnotesize Each $(r,s)$ box
    contains the dimension of the highest-weight vectors of
    $\repX^{+}_{r,s}$ and $\repX^{-}_{r,s}$, in this order.  The
    $(p_+-1)\times(p_--1)=2\times1$ subtable also contains the
    dimensions of $\repX_{r,s}$ (which are $\repX_{1,1}$ in this
    case), shown in parentheses.  The (infinite) Virasoro content
    follows from the decompositions in~\bref{prop:x-vir}.}
  \begin{tabular}{|c|c|c|}
    \hline
    $\ffrac{5}{8}$, \ $\ffrac{33}{8}$&
    $\ffrac{1}{8}$, \ $\ffrac{21}{8}$&
    $-\ffrac{1}{24}$, \ $\ffrac{35}{24}$
    \rule[-10pt]{0pt}{26pt}
    \\
    \hline
    $2$, \ $7$ \ $(0)$&$1$, \ $5$ \ $(0)$&
    $\ffrac{1}{3}$, \ $\ffrac{10}{3}$\rule[-10pt]{0pt}{26pt}
    \\
    \hline
  \end{tabular}
  \label{table-kac}
\end{table}

\subsection{``Verma'' modules}\label{subsec:Verma}
The $\talgL\!$-modules $\repV_{r,s}^\pm$ (see~\eqref{eq:w-verma-def})
are $\talgW\!$-modules (simply because $\talgW$ is a subalgebra
in~$\talgL$).  In referring to the modules $\repV_{r,s}^\pm$ as
$\talgW$-modules, it is convenient to call them the Verma modules of
the $\talgW$ algebra. (Their counterparts for the
Kazhdan--Lusztig-dual quantum group~\cite{[FGST-q]} are indeed Verma
modules, but investigation of the Verma properties of
$\repV_{r,s}^\pm$ is a separate problem, which we do not consider here
and only use the convenient and suggestive name for these modules.)
We now describe their subquotients.

\begin{prop}\label{prop:w-verma-struct}
  The subquotient structure of the $\repV_{r,s}^{\pm}$ is as follows.
  
  \begin{enumerate}
  \item $\repV_{p_+,p_-}^{\pm}\simeq\repX_{p_+,p_-}^{\pm}$.
    
  \item For $1\leq r\leq p_+\,{-}\,1$, $\repX_{p_+-r,p_-}^\mp \subset
    \repV_{r,p_-}^{\pm}$ and
    $\repV_{r,p_-}^{\pm}/\repX_{p_+-r,p_-}^\mp \simeq
    \repX_{r,p_-}^\pm$.
    
  \item For $1\leq s\leq p_-\,{-}\,1$, $\repX_{p_+,p_--s}^\mp \subset
    \repV_{p_+,s}^{\pm}$ and
    $\repV_{p_+,s}^{\pm}/\repX_{p_+,p_--s}^\mp \simeq
    \repX_{p_+,s}^\pm$.
    
  \item For $1\leq r\leq p_+\,{-}\,1$ and $1\leq s\leq p_-\,{-}\,1$, \ 
    $\repV_{r,s}^{+}$ admits a filtration
    \begin{equation*}
      \rep{H}_0\subset\rep{H}_1\subset\repV_{r,s}^{+},
    \end{equation*}
    where $\rep{H}_0\simeq\repX_{p_+-r,p_--s}^+$,
    $\rep{H}_1/\rep{H}_0\simeq\repX_{p_+-r,s}^-
    \oplus\repX_{r,s}\oplus\repX_{r,p_--s}^-$, and
    $\repV_{r,s}^{+}/\rep{H}_1\simeq \repX_{r,s}^+$; and
    $\repV_{r,s}^{-}$ admits a filtration
    \begin{equation*}
      \rep{H}_0\subset\rep{H}_1\subset\repV_{r,s}^{-},
    \end{equation*}
    where $\rep{H}_0\simeq\repX_{p_+-r,p_--s}^-$,
    $\rep{H}_1/\rep{H}_0\simeq\repX_{p_+-r,s}^+
    \oplus\repX_{r,p_--s}^+$, and
    $\repV_{r,s}^{+}/\rep{H}_1\simeq\repX_{r,s}^-$.
  \end{enumerate}
\end{prop}

The subqotient structure is clear from Figs.~\ref{fig:W-action},
\ref{fig:W-action-min}, \ref{fig:W-action-r,p_--min},
and~\ref{fig:W-action-r,p_-}.  For example, we consider
Fig.~\ref{fig:W-action}.  The filled dots {\small $\bullet$}
constitute the subquotient~$\repX_{r,s}^{+}$, the cross {\small
  $\times$} is $\repX_{r,s}$, the {\small $\circ$} and {\small
  $\scrBox$} are combined into~$\repX_{r,p_--s}^{-}$
and~$\repX_{p_+-r,s}^{-}$ respectively, and the
{\small$\blacktriangle$} are combined into~$\repX_{p_+-r,p_--s}^{+}$.

\subsection{Bimodule structure of $\bigrepF$}\label{subsec:bimodule}
The quantum group~$\tqalgA$ acts in the space $\bigrepF$, which is in
fact a $(\talgW,\tqalgA)$-bimodule.

The subquotient structure of the \textit{bimodule} $\bigrepF$ shows
the origin of the multiplicities in decomposition~\eqref{eq:FF-dec}:
each indecomposable $\talgW$-module enters in several copies produced
by the action of~$\qalgB$.  

\begin{prop}\mbox{}
  As a $(\talgW,\tqalgA)$-bimodule, the space $\bigrepF$ decomposes as
  \begin{equation}\label{eq:verma-bi}
    \bigrepF=\bigoplus_{(r,s)\in\seti}\rep{Q}_{r,s}
  \end{equation}
  where $\rep{Q}_{r,s}$ are indecomposable
  $(\talgW,\tqalgA)$-bimodules with the following structure.
  \begin{enumerate}
  \item If $(r,s)=(0,0)$, then
    $\rep{Q}_{0,0}=\repX^+_{p_+,p_-}\boxtimes\XX^{+}_{p_+,p_-}$\textup;
    
  \item if $(r,s)=(p_+,0)$, then
    $\rep{Q}_{p_+,0}=\repX^-_{p_+,p_-}\boxtimes\XX^{-}_{p_+,p_-}$\textup;
    
  \item if $1\leq r\leq p_+\,{-}\,1$ and $s=0$, then $\rep{Q}_{r,0}$
    has the subquotient structure described in Fig.~\ref{fig:Qr0}\textup;
    \begin{figure}[tbp]
      \begin{center}
        \xymatrix@=34pt{
          &*{}&{\repX^+_{r,p_-}\makebox[0pt][l]{$\boxtimes\,
              \XX^{-}_{p_+-r,p_-}$}}
          \ar[d] &\ar@{-->}[drrr]+<25pt,10pt>_(0.6){\ep}&*{}&*{}
          &{\repX^-_{p_+-r,p_-}\makebox[0pt][l]{$\boxtimes\,
              \XX^{+}_{r,p_-}$}}
          \ar[d]\ar@{-->}[dllll]+<40pt,10pt>^(0.6){\ep}&\\
          &*{}&{\repX^-_{p_+-r,p_-}\makebox[0pt][l]{$\boxtimes\,
              \XX^{-}_{p_+-r,p_-}$}}
          &&*{}&*{}&{\repX^+_{r,p_-}\makebox[0pt][l]{$\boxtimes\,
              \XX^{+}_{r,p_-}$}}&}
        \caption[Subquotient structure of
        $\rep{Q}_{r,0}$]{\captionfont{Subquotient structure of
            $\rep{Q}_{r,0}$}.  \footnotesize $\boxtimes$ denotes
          the external tensor product.
           Solid lines denote the
          $\talgW$-action and dashed lines denote the
          $\tqalgA$-action.  }
        \label{fig:Qr0}
      \end{center}
    \end{figure}
    
  \item if $r=0$ and $1\leq s\leq p_-\,{-}\,1)$, then $\rep{Q}_{0,s}$
    has the subquotient structure obtained from Fig.~\ref{fig:Qr0} by
    changing $\repX^+_{r,p_-}\to\repX^+_{p_+,s}$,
    $\repX^+_{p_+-r,p_-}\to\repX^+_{p_+,p_--s}$,
    $\XX^+_{r,p_-}\to\XX^+_{p_+,s}$,
    $\XX^+_{p_+-r,p_-}\to\XX^+_{p_+,p_--s}$, and~$\ep\to\fm$\textup;
    
  \item if $(r,s)\in\setii$, then $\rep{Q}_{r,s}$ has the subquotient
    structure described in Fig.~\ref{fig:Qrs}.
    \begin{figure}[tbhp]
      \begin{center}
        \includegraphics[bb=1.4in 6.8in 8in 10.3in,
        clip]{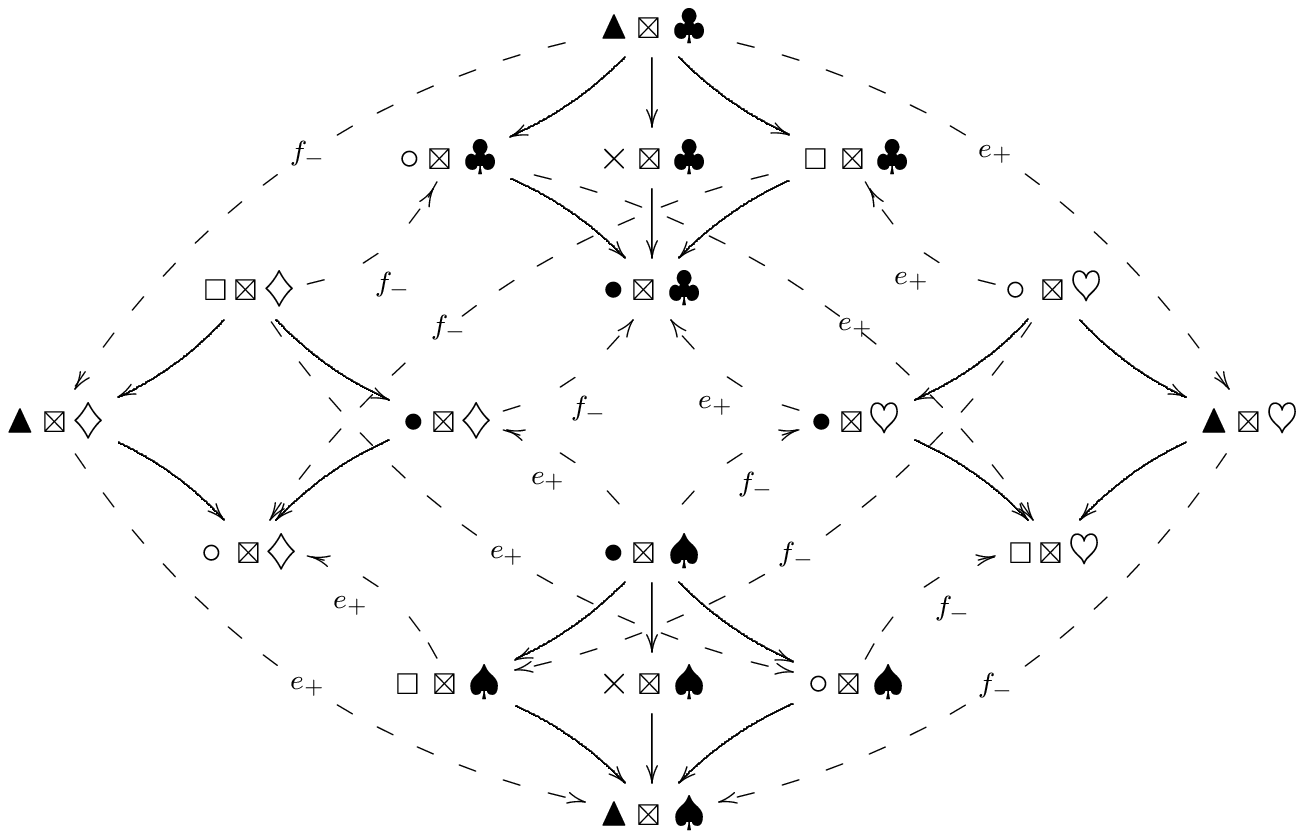}
\caption[Subquotient structure of
        $\rep{Q}_{r,s}$]{\captionfont{Subquotient structure of
            $\rep{Q}_{r,s}$}.  \footnotesize We use the notation
          \begin{alignat*}{4}
            &\bullet=\repX^{+}_{r,s},&\quad
            &\circ=\repX^{-}_{r,p_--s}&,\quad
            &\scrBox=\repX^{-}_{p_+-r,s}&,\quad
            &\blacktriangle=\repX^{+}_{p_+-r,p_--s},\\
            \intertext{and $\times=\repX_{r,s}$ for $\talgW$-modules,
              and}
            &\clubsuit=\XX^{+}_{r,s},&
            &\diamondsuit=\XX^{-}_{r,p_--s},&\quad
            &\heartsuit=\XX^{-}_{p_+-r,s},&
            &\spadesuit=\XX^{+}_{p_+-r,p_--s}
          \end{alignat*}
          for quantum-group modules.  Solid lines show the action of
          $\algW_{p_+,p_-}$-generators and dashed lines show the
          action of $\tqalgA$-generators, and $\boxtimes$ denotes
          external tensor product.}
        \label{fig:Qrs}
      \end{center}
    \end{figure}
  \end{enumerate}
\end{prop}

\subsubsection{}In particular, forgetting the quantum-group
structure, we obtain a decomposition of $\bigrepF$ into a finite sum
of $\talgW$ Verma modules with multiplicities (dimensions of
irreducible $\tqalgA$-modules) as
\begin{multline}
  \label{eq:FF-dec}
  \bigrepF=\bigoplus_{r=1}^{p_+-1} \bigoplus_{s=1}^{p_--1}
  (p_+\!-\!r)(p_-\!-\!s)\bigl(\repV_{r,s}^{+}\oplus
  \repV_{r,s}^{-}\bigr)
  \oplus\bigoplus_{r=1}^{p_+-1}
  (p_+\!-\!r)p_-\bigl(
  \repV_{r,p_-}^{+}\oplus\repV_{r,p_-}^{-}\bigr)\\
  \oplus \bigoplus_{s=1}^{p_--1}
  (p_-\!-\!s)p_+\bigl(\repV_{p_+,s}^{+}\oplus\repV_{p_+,s}^{-}\bigr)
  \oplus
  p_+p_-\bigl(\repV_{p_+,p_-}^{+}\oplus\repV_{p_+,p_-}^{-}\bigr).
\end{multline}
The structure of $\bigrepF$ as a $(\talgW,\tqalgA)$-bimodule can be
taken as a starting point for constructing the projective
$\talgW$-modules by the method in~\cite{[FFHST]}.

\section{The space of torus amplitudes}\label{sec:torus}
In this section, we find the $\SLiiZ$-representation generated from
the space of characters~$\Grring$ in the logarithmic $(p_+,p_-)$
model.  With Conjecture~\bref{conj:gch} assumed, this gives the space
of torus amplitudes.  In~\bref{subsec:char}, we calculate the
characters of irreducible $\talgW$-modules in terms of
theta-constants, see~\eqref{theta-const}.  It is then straightforward
to calculate modular transformations of the characters, which we do
in~\bref{subsec:gch}.  This extends the space of characters by
$\oC[\tau]/\tau^3$, giving rise to the space of generalized characters
$\sgch$ (which, as noted above, most probably coincides with the space
$\TA$ of torus amplitudes). (Theta-function derivatives enter the
characters through the second order; this can be considered a
``technical'' reason for the explicit occurrences of~$\tau$
and~$\tau^2$ in the modular transformation properties.) \ The analysis
of the $\SLiiZ$-representation on~$\sgch$ then yields the results
in~\bref{thm:R-decomp}.  In~\bref{sec:mod-inv}, we use the established
decomposition of the $\SLiiZ$ action to briefly consider modular
invariants.

\subsection{$\smash{\talgW}$-characters}\label{subsec:char}
Let
\begin{alignat*}{2}
  \chi^{\pm}_{r,s}(q)&=\Tr_{\repX^{\pm}_{r,s}}q^{L_0-\frac{c}{24}},&
  \qquad &1\leq r\leq p_+,\quad 1\leq s\leq p_-,
  \\
  \chi_{r,s}(q)&=\Tr_{\repX_{r,s}}q^{L_0-\frac{c}{24}},&
  \quad  &(r,s)\in\setii,
\end{alignat*}
with $c$ given by~\eqref{eq:centr-charge}, be the characters of the
irreducible $\talgW$-representations $\repX^{\pm}_{r,s}$ and
$\repX_{r,s}$ (see~\bref{subsec:w-mod}).

\begin{prop}\label{find-chars}
  The irreducible $\talgW$-representation characters are given by
  \begin{align}
    \label{prop-chi_rs}
    \chi_{r,s}={}& \ffrac{1}{\eta}
    (\theta_{p_+s-p_-r}-\theta_{p_+s+p_-r}),\quad
    (r,s)\in\setii,\\
    \label{prop-chi+_rs}
    \chi^+_{r,s}={}&
    \ffrac{1}{(p_+p_-)^2\eta}
    \Bigl(\theta''_{p_+s+p_-r}-\theta''_{p_+s-p_-r}\\*
    {}& -(p_+s\!+\!p_-r)\theta'_{p_+s+p_-r}
    +(p_+s\!-\!p_-r)\theta'_{p_+s-p_-r}\notag\\*
    {}&+\ffrac{(p_+s\!+\!p_-r)^2}{4}\,\theta_{p_+s+p_-r}
    -\ffrac{(p_+s\!-\!p_-r)^2}{4}\,\theta_{p_+s-p_-r}\Bigr),\quad
    \kern-10pt 1\leq  r\leq p_+,\quad \kern-10pt
    1\leq s\leq p_-,\kern-15pt \notag\\
    \label{prop-chi-_rs}
    \chi^-_{r,s}={}& \ffrac{1}{(p_+p_-)^2\eta}\Bigl(
    \theta''_{p_+p_--p_+s-p_-r}- \theta''_{p_+p_-+p_+s-p_-r}\\
    {}& +(p_+s\!+\!p_-r)\theta'_{p_+p_--p_+s-p_-r}
    +(p_+s\!-\!p_-r)\theta'_{p_+p_-+p_+s-p_-r}\notag\\
    {}& +\ffrac{(p_+s+p_-r)^2\!-\!(p_+p_-)^2}{4}\,
    \theta_{p_+p_--p_+s-p_-r}\notag\\
    {}&-\ffrac{(p_+s-p_-r)^2\!-\!(p_+p_-)^2}{4}\,
    \theta_{p_+p_-+p_+s-p_-r}
    \Bigr),\quad1\leq r\leq p_+,\quad 1\leq s\leq p_-.\notag
  \end{align}
\end{prop}
\begin{proof}
  We first recall the well-known irreducible Virasoro
  characters~\cite{[RC]}
  \begin{multline}\label{eq:rocha-carr}
    \mathrm{char}\,\repJ_{r,s;n}(q)
    =\ffrac{q^{\frac{1-c}{24}}}{\eta(q)}\Bigl(
    \sum_{m\geq0}q^{\Delta_{r,s;n+2m}}
    +\sum_{m\geq1}q^{\Delta_{r,s;-n-2m}}\\
    -\sum_{m\geq0}q^{\Delta_{r,p_--s;n+2m+1}}
    -\sum_{m\geq0}q^{\Delta_{r,p_--s;-n-2m-1}}\Bigr).
  \end{multline}
  In particular, for the characters of $\repX_{r,s}$, we immediately
  have
  \begin{equation*}
    \mathrm{char}\,\repX_{r,s}(q)
    =\ffrac{q^{\frac{1-c}{24}}}{\eta(q)}\sum_{m\in\oZ}
    (q^{\Delta_{r,s;2m}}-q^{\Delta_{r,p_--s;2m+1}}),
  \end{equation*}
  which gives~\eqref{prop-chi_rs} when rewritten in terms of the
  theta-constants.
  
  Next, from~\bref{prop:x-vir}, we have that the character of each
  $\repX^+_{r,s}$ for $1\leq r\leq p_+$ and $1\leq s\leq p_-$ is given
  by
  \begin{equation*}
    \chi^+_{r,s}(q)=\Tr_{\repX^+_{r,s}}q^{L_0-\frac{c}{24}}
    =\sum_{a\geq0}(2a+1)\,\mathrm{char}\,\repJ_{r,p_--s;2a+1}(q).
  \end{equation*}
  Substituting~\eqref{eq:rocha-carr} here, we have
  \begin{equation*}
    \chi^+_{r,s}(q)=\Sigma_1+\Sigma_2,
  \end{equation*}
  where
  \begin{align*}
    \Sigma_1&=\sum_{a\geq0}(2a+1)q^{\frac{1-c}{24}}
    \Bigl(\sum_{m\geq0}q^{\Delta_{r,p_--s;2a+2m+1}}
    +\sum_{m\geq1}q^{\Delta_{r,p_--s;-2a-2m-1}}\Bigr),\\
    \Sigma_2&=-\sum_{a\geq0}(2a+1)q^{\frac{1-c}{24}}
    \Bigl(\sum_{m\geq0}q^{\Delta_{r,s;2a+2m+2}}
    +\sum_{m\geq0}q^{\Delta_{r,s;-2a-2m-2}}\Bigr).
  \end{align*}
  Setting $\bar\Delta_{r,s;n}=\Delta_{r,s;n}+\frac{1-c}{24}$ for
  brevity, we obtain
  \begin{align*}
    \Sigma_1={}&
    \sum_{m\geq0}\sum_{a=0}^{m}(2a+1)q^{\bar\Delta_{r,p_--s;2m+1}}
    +\sum_{m\leq-2}\sum_{a=0}^{-m-2}(2a+1)q^{\bar\Delta_{r,p_--s;2m+1}}\\
    ={}&\sum_{m\in\oZ}(m+1)^2q^{\bar\Delta_{r,p_--s;2m+1}}=
    \sum_{m\in\oZ}m^2q^{p_+p_-(m-\frac{p_+s+p_-r}{2p_+p_-})}\\
    ={}&
   \sum_{m\in\oZ}
    \bigl(m-\ffrac{p_+s+p_-r}{2p_+p_-}\bigr)^2
    q^{p_+p_-(m-\frac{p_+s+p_-r}{2p_+p_-})^2}+{}\\
    &\qquad{}
    +\ffrac{p_+s+p_-r}{p_+p_-}\sum_{m\in\oZ}
    \bigl(m-\ffrac{p_+s+p_-r}{2p_+p_-}\bigr)
    q^{p_+p_-(m-\frac{p_+s+p_-r}{2p_+p_-})^2}\\
    &\qquad{}
    +\ffrac{(p_+s+p_-r)^2}{(2p_+p_-)^2}\sum_{m\in\oZ}
    q^{p_+p_-(m-\frac{p_+s+p_-r}{2p_+p_-})^2}\\
    ={}&
    \ffrac{1}{(p_+p_-)^2}\,
    \theta''_{-p_+s-p_-r}
    + \ffrac{p_+s\!+\!p_-r}{(p_+p_-)^2}\,\theta'_{-p_+s-p_-r}
    +\ffrac{(p_+s\!+\!p_-r)^2}{(2p_+p_-)^2}\,
    \theta_{-p_+s-p_-r},
  \end{align*}
  and a similar calculation gives
  \begin{equation*}
    \Sigma_2=    
    -\ffrac{1}{(p_+p_-)^2}\theta''_{p_+s-p_-r}
    +\ffrac{p_+s-p_-r}{(p_+p_-)^2}\theta'_{p_+s-p_-r}
    -\ffrac{(p_+s-p_-r)^2}{(2p_+p_-)^2}\,\theta_{p_+s-p_-r},
  \end{equation*}
  whence~\eqref{prop-chi+_rs} follows.  Similar calculations give the
  character in~\eqref{prop-chi-_rs}.
\end{proof}

\begin{rem}
  {}From the definition of the Verma modules $\repV_{r,s}^{\pm}$ in
  Eqs.~\eqref{eq:w-verma-def}, we calculate their characters as
  \begin{align*}
    \mathrm{char}\,\repV_{r,s}^{+}(q)&=
    \ffrac{q^{\frac{1-c}{24}}}{\eta(q)}\sum_{n\in\oZ}q^{\Delta_{r,s,2n}}
    =\ffrac{1}{\eta(q)}\,\theta_{p_+s-p_-r,\,p_+p_-}(q),\\
    \mathrm{char}\,\repV_{r,s}^{-}(q)&=
    \ffrac{q^{\frac{1-c}{24}}}{\eta(q)}
    \sum_{n\in\oZ}q^{\Delta_{r,s,2n+1}}
    =\ffrac{1}{\eta(q)}\,\theta_{p_+p_-+p_+s-p_-r,\,p_+p_-}(q).
  \end{align*}
  On the other hand, from the subquotient structure of Verma modules
  described in~\bref{prop:w-verma-struct}, we obtain the identities
  \begin{align*}
    \mathrm{char}\,\repV_{r,s}^{+}(q)&=\chi_{r,s}(q)+
    \chi^+_{r,s}(q)+\chi^-_{r,p_--s}(q)
    +\chi^-_{p_+-r,s}(q)+\chi^+_{p_+-r,p_--s}(q),\\
    \mathrm{char}\,\repV_{r,s}^{-}(q)
    &=\chi^-_{r,s}(q)+\chi^+_{r,p_--s}(q)
    +\chi^+_{p_+-r,s}(q)+\chi^-_{p_+-r,p_--s}(q)
  \end{align*}
  for $(r,s)\in\setii$, which shows that, as could be expected, the
  space of Verma-module characters coincides with the space of
  theta-functions (\textit{not} including their derivatives).  We also
  note that similar expressions for characters involving second
  derivatives of theta-functions were proposed in~\cite{[N]}.
\end{rem}

\subsection{The space $\smash{\sgch}$}\label{subsec:gch}
We now construct the $\SLiiZ$-representation $\sgch$ generated from
the space $\Grring$ of the irreducible-representation characters and
simultaneously obtain the decomposition in~\bref{thm:R-decomp}.

\subsubsection{}\label{sec:chi2rho}
First, $\Grring$ contains the subspace~$R_{\mathrm{min}}$ spanned by
$\chi_{r,s}$ with $(r,s)\in\setii$\,---\,the Virasoro minimal-model
characters, which evidently carry an $\SLiiZ$-representation.  Second,
as we see shortly, the $\half(p_+\,{+}\,1)(p_-\,{+}\,1)$-dimensional
space~$R_{\mathrm{proj}}\subset\Grring$ linearly spanned by the
functions
\begin{alignat}{2}
  \wkappap_{r,s} ={}&\chi_{r,s}+ 2\chi^+_{r,s}+2\chi^-_{r,p_--s}
  +2\chi^-_{p_+-r,s}+2\chi^+_{p_+-r,p_--s},&\quad&(r,s)\in\setii,
  \notag\\
  \wkappap_{0,s}={}&
  2\chi^+_{p_+,p_--s}+2\chi^-_{p_+,s},&&1\leq s\leq p_-\,{-}\,1,
  \notag\\
  \wkappap_{r,0}={}&
  2\chi^+_{p_+-r,p_-}+2\chi^-_{r,p_-},&&1\leq r\leq p_+\,{-}\,1,
  \label{the-kappa}\\
  \wkappap_{0,0}={}&2\chi^+_{p_+,p_-},&
  \notag\\
  \wkappap_{p_+,0}={}&2\chi^-_{p_+,p_-}&
  \notag
\end{alignat}
(which can be identified with the characters of projective
$\talgW$-modules) also carries an $\SLiiZ$-representation.

Next, the $(p_+p_-\,{-}\,1)$-dimensional complement of
$R_{\mathrm{min}}\oplus R_{\mathrm{proj}}$ in $\Grring$ contains no
more subspaces closed under the $\SLiiZ$-action.  It is convenient to
choose the functions
\begin{alignat*}{2}
  \wrhom_{r,s} ={}&\ffrac{p_+s-p_-r}{2}\chi_{r,s}
  - p_+(p_-\!-\!s)(\chi^+_{r,s}
  +\chi^-_{p_+-r,s})&\\
  &{} +p_+s(\chi^+_{p_+-r,p_--s}
  +\chi^-_{r,p_--s}),&\quad&(r,s)\in\setii,\\
  \wrhom_{0,s}={}&
   p_+(s\chi^+_{p_+,p_--s}-(p_-\!-\!s)\chi^-_{p_+,s}),
  &&1\leq s\leq p_-\!-\!1,\\
  \wrhop_{r,s} ={}&\ffrac{p_-r-p_+s}{2}\chi_{r,s}
  +p_-(r-p_+)(\chi^+_{r,s}
  +\chi^-_{r,p_--s})&\\
  &{} +p_-r(\chi^+_{p_+-r,p_--s}
  +\chi^-_{p_+-r,s}),&&(r,s)\in\setii,\\
  \wrhop_{r,0}={}&
  p_-(r\chi^+_{p_+-r,p_-}-(p_+\!-\!r)\chi^-_{r,p_-}),
  &&1\leq r\leq p_+\!-\!1,\\
  \wxi_{r,s}
  ={}&p_+p_-\Bigl((p_+\!-\!r)(p_-\!-\!s)
  \chi^+_{r,s}+rs\chi^+_{p_+-r,p_--s}
  -\ffrac{(p_+s\!-\!p_-r)^2}{4p_+p_-}\chi_{r,s}\kern-70pt&\\
  &{}-(p_+\!-\!r)s\chi^-_{r,p_--s} -r(p_-\!-\!s)\chi^-_{p_+-r,s}
  \Bigr),
  &&(r,s)\in\setii
\end{alignat*}
as a basis in this complement.

We then define
\begin{equation}\label{varphi-def}
  \begin{alignedat}{2}
    \wphip_{r,s}(\tau)={}&\tau\wrhop_{r,s}(\tau),
    &\qquad &(r,s)\in\setp,\\
    \wphim_{r,s}(\tau)={}&\tau\wrhom_{r,s}(\tau),& &(r,s)\in\setm,\\
    \wpsi_{r,s}(\tau)
    ={}&2\tau\wxi_{r,s}(\tau)+i\pi p_+p_-\wkappam_{r,s}(\tau),&\qquad
    &(r,s)\in\setii,\\  
    \wzeta_{r,s}(\tau) ={}&\tau^2\wxi_{r,s}(\tau)
    +i\pi p_+p_-\tau\wkappam_{r,s}(\tau),&
    &(r,s)\in\setii.
  \end{alignedat}
\end{equation}
\begin{prop}\label{prp:closed}
  The $\SLiiZ$-representation~$\sgch$ generated from $\Grring$ is the
  linear span of the $\half(3p_+\,{-}\,1)(3p_-\,{-}\,1)$ functions
  \begin{itemize}
  \item[] $\wkappam_{r,s}(\tau)$, $\wxi_{r,s}(\tau)$,
    $\wpsi_{r,s}(\tau)$, and $\wzeta_{r,s}(\tau)$ with
    $(r,s)\in\setii$,
    
  \item[] $\wkappap_{r,s}(\tau)$ with $(r,s)\in\seti$,
    
  \item[] $\wrhop_{r,s}(\tau)$ and $\wphip_{r,s}(\tau)$ with
    $(r,s)\in\setp$, and
    
  \item[] $\wrhom_{r,s}(\tau)$ and $\wphim_{r,s}(\tau)$ with
    $(r,s)\in\setm$.
  \end{itemize}
\end{prop}
This is summarized in Table~\ref{table1}, where we also indicate the
$\SLiiZ$-representation and its dimension spanned by each group of
functions.
\begin{table}[tbph]
  \centering

  \medskip

  \caption{\captionfont{The basis in $\sgch$}}\label{table1}
  
  \begin{tabular}{|p{3cm}|p{4cm}|p{5cm}|}
    \hline
    subrep.&dimension&basis\\
    \hline
    $R_{\mathrm{min}}$&
    $\half(p_+\,{-}\,1)(p_-\,{-}\,1)$&$\wkappam_{r,s}$,
    $(r,s)\in\setii$\\
    \hline
    $R_{\mathrm{proj}}$&
    $\half(p_+\,{+}\,1)(p_-\,{+}\,1)$&$\wkappap_{r,s}$,
    $(r,s)\in\seti$\\
    \hline
    $\oC^2\tensor R_{\boxslash}$&
    $2\cdot \half(p_+\,{-}\,1)(p_-\,{+}\,1)$&
    $\wrhop_{r,s}$, $\wphip_{r,s}$,  $(r,s)\in\setp$\\
    \hline
    $\oC^2\tensor R_{\boxbslash}$&
    $2\cdot \half(p_+\,{+}\,1)(p_-\,{-}\,1)$&
    $\wrhom_{r,s}$, $\wphim_{r,s}$, $(r,s)\in\setm$\\
    \hline
    $\oC^3\tensor R_{\mathrm{min}}$&$
    3\cdot \half(p_+\,{-}\,1)(p_-\,{-}\,1)$&
    $\wxi_{r,s}$, $\wpsi_{r,s}$, $\wzeta_{r,s}$, $(r,s)\in\setii$\\
    \hline
  \end{tabular}
\end{table}

The proof amounts to the following two lemmas, the first of which
shows that $\sgch$ is closed under the $\modS$-transformation and the
second that it is closed under the $\modT$-transfor\-mation.
\begin{lemma}\label{S-lemma}
  \begin{align}
    \label{eq:S-kappam}
    \wkappam_{r,s}(-\ffrac{1}{\tau})=
    {}&-\ffrac{2\sqrt{2}}{\sqrt{\mathstrut p_+p_-}}
    \sum_{(r',s')\in\setii}\!\!
    (-1)^{rs'+r's}\sin\ffrac{\pi p_- rr'}{p_+}
    \,\sin\!\ffrac{\pi p_+ ss'}{p_-}\,\wkappam_{r',s'}(\tau),\\*
    &\kern200pt(r,s)\in\setii,\notag\\
    \label{eq:S-kappap}
    \wkappap_{r,s}(-\ffrac{1}{\tau})=
    {}&\ffrac{\sqrt{2}}{\sqrt{\mathstrut p_+p_-}}
    \Bigl(\sum_{(r',s')\in\setii}\!\!
    2(-1)^{rs'+r's}\cos\ffrac{\pi p_- rr'}{p_+}
    \,\cos\ffrac{\pi p_+ ss'}{p_-}\,\wkappap_{r',s'}(\tau)\\
    &+\!\sum_{r'=1}^{p_+-1}(-1)^{r's}\cos\ffrac{\pi p_- rr'}{p_+}
    \,\wkappap_{r',0}(\tau)
    +\!\sum_{s'=1}^{p_--1}(-1)^{rs'}
    \cos\ffrac{\pi p_+ ss'}{p_-} \,\wkappap_{0,s'}(\tau)\notag\\
    &+\half\,\wkappap_{0,0}(\tau)+\half(-1)^{p_-r+p_+s}
    \,\wkappap_{p_+,0}(\tau)\Bigr),
    \quad(r,s)\in\seti,\notag\\
    \label{eq:S-rhom}
    \wrhom_{r,s}(-\ffrac{1}{\tau})=
    {}&-i\ffrac{\sqrt{2}}{\sqrt{\mathstrut p_+p_-}}\Bigl(
    \sum_{(r',s')\in\setii}\!\!
    2 (-1)^{rs'+r's}\sin\ffrac{\pi p_+ ss'}{p_-}
    \,\cos\ffrac{\pi p_- rr'}{p_+}
    \,\wphim_{r',s'}(\tau)\\
    &+\sum_{s'=1}^{p_- - 1}(-1)^{rs'}
    \,\sin\ffrac{\pi p_+ ss'}{p_-}
    \,\wphim_{0,s'}(\tau)\Bigr),
    \quad(r,s)\in\setm,\notag\\
    \label{eq:S-phim}
    \wphim_{r,s}(-\ffrac{1}{\tau})=
    {}&i\ffrac{\sqrt{2}}{\sqrt{\mathstrut p_+p_-}}\Bigl(
    \sum_{(r',s')\in\setii}\!\!
    2 (-1)^{rs'+r's}\sin\ffrac{\pi p_+ ss'}{p_-}
    \,\cos\ffrac{\pi p_- rr'}{p_+}\,\wrhom_{r',s'}(\tau)\\
    &+\sum_{s'=1}^{p_- - 1}(-1)^{rs'}
    \sin\ffrac{\pi p_+ ss'}{p_-}\,\wrhom_{0,s'}(\tau)\Bigr),
    \quad(r,s)\in\setm,\notag\\
    \label{eq:S-rhop}
    \wrhop_{r,s}(-\ffrac{1}{\tau})=
    {}&-i\ffrac{\sqrt{2}}{\sqrt{\mathstrut p_+p_-}}\Bigl(
    \sum_{(r',s')\in\setii}\!\!
    2 (-1)^{rs'+r's}\sin\ffrac{\pi p_- rr'}{p_+}
    \,\cos\ffrac{\pi p_+ ss'}{p_-}\,\wphip_{r',s'}(\tau)\\
    &+\sum_{r'=1}^{p_+ - 1}(-1)^{r's}\sin\ffrac{\pi p_- rr'}{p_+}
    \,\wphip_{r',0}(\tau)\Bigr),
    \quad(r,s)\in\setp,\notag\\
    \label{eq:S-phip}
    \wphip_{r,s}(-\ffrac{1}{\tau})=
    {}&i\ffrac{\sqrt{2}}{\sqrt{\mathstrut p_+p_-}}\Bigl(
    \sum_{(r',s')\in\setii}\!\!
    2 (-1)^{rs'+r's}\sin\ffrac{\pi p_- rr'}{p_+}
    \,\cos\ffrac{\pi p_+ ss'}{p_-}\,\wrhop_{r',s'}(\tau)\\
    &+\sum_{r'=1}^{p_+ - 1}(-1)^{r's}\sin\ffrac{\pi p_- rr'}{p_+}
    \,\wrhop_{r',0}(\tau)\Bigr),
    \quad(r,s)\in\setp,\notag\\
    \label{eq:S-xi}
    \wxi_{r,s}(-\ffrac{1}{\tau})=
    {}&{-}\ffrac{2\sqrt{2}}{\sqrt{\mathstrut p_+p_-}}
    \sum_{(r',s')\in\setii}\!\!
    (-1)^{rs'+r's}\sin\ffrac{\pi p_- rr'}{p_+}
    \,\sin\ffrac{\pi p_+ ss'}{p_-}\,\wzeta_{r',s'}(\tau),\\
    &\kern200pt(r,s)\in\setii,\notag\\
    \label{eq:S-psi}
    \wpsi_{r,s}(-\ffrac{1}{\tau})=
    {}&\ffrac{2\sqrt{2}}{\sqrt{\mathstrut p_+p_-}}
    \sum_{(r',s')\in\setii}\!\!
    (-1)^{rs'+r's}\sin\ffrac{\pi p_- rr'}{p_+}
    \,\sin\ffrac{\pi p_+ ss'}{p_-}\,\wpsi_{r',s'}(\tau),\\
    &\kern200pt(r,s)\in\setii,\notag\\
    \label{eq:S-zeta}
    \wzeta_{r,s}(-\ffrac{1}{\tau})=
    {}&{-}\ffrac{2\sqrt{2}}{\sqrt{\mathstrut p_+p_-}}
    \sum_{(r',s')\in\setii}\!\!
    (-1)^{rs'+r's}\sin\ffrac{\pi p_- rr'}{p_+}\,
    \sin\ffrac{\pi p_+ ss'}{p_-}\,\wxi_{r',s'}(\tau),\\
    &\kern200pt(r,s)\in\setii.\notag
  \end{align}
\end{lemma}
\begin{proof}
  The standard modular transformation formulas for the theta-constants
  are
  \begin{align*}
    \theta_{s,p}(-\ffrac{1}{\tau})={}&\sqrt{\ffrac{-i\tau}{2p}}
    \sum_{r=0}^{2p-1} e^{-i\pi\frac{rs}{p}}\,\theta_{r,p}(\tau),\\
    \theta'_{s,p}(-\ffrac{1}{\tau})={}&\tau\sqrt{\ffrac{-i\tau}{2p}}
    \sum_{r=0}^{2p-1} e^{-i\pi\frac{rs}{p}}\theta'_{r,p}(\tau),\\
    \theta''_{s,p}(-\ffrac{1}{\tau})={}&\sqrt{\ffrac{-i\tau}{2p}}
    \sum_{r=0}^{2p-1}e^{-i\pi\frac{rs}{p}}\bigl( \tau^2
    \theta''_{r,p}(\tau) + i\pi p\tau\theta_{r,p}(\tau)\bigr)
  \end{align*}
  and the eta function transforms as
  \begin{gather*}
    \eta(\tau+1)=e^{\frac{i\pi}{12}}\eta(\tau),\qquad
    \eta(-\ffrac{1}{\tau})= \sqrt{-i\tau}\,\eta(\tau).
  \end{gather*}  
  
  We next use the resummation formula~\eqref{eq:the-sum}.
  Substituting
  \begin{equation*}
    g(r)=e^{-i\pi\frac{rs}{p_+p_-}},
    \quad
    h_+(r)=\ffrac{1}{\sqrt{\mathstrut 2p_+p_-}}
    \ffrac{\theta_r(\tau)}{\eta(\tau)},
    \quad
    u^+_+(r,s)=\wkappap_{r,s},
    \quad
    u^-_+(r,s)=\wkappam_{r,s}
  \end{equation*}
  there, we find the transformation of the theta-constants as
  \begin{multline*}
    \frac{\theta_{s}(-\frac{1}{\tau})}{\eta(-\ffrac{1}{\tau})}
    =\ffrac{1}{\sqrt{\mathstrut 2p_+p_-}} 
    \ffrac{1}{\eta(\tau)}
    \sum_{r=0}^{2p_+p_--1} e^{-i\pi\frac{rs}{p_+p_-}}\theta_{r}(\tau)\\
    = \ffrac{1}{\sqrt{\mathstrut
        2p_+p_-}}
    \biggl(\half\,\wkappap_{0,0}(\tau)
    + \ffrac{(-1)^s\!}{2}\,\wkappap_{p_+,0}(\tau)
    + \sum_{r'=1}^{p_+-1}\cos\ffrac{\pi r's}{p_+}\,\wkappap_{r',0}(\tau)
    + \sum_{s'=1}^{p_--1}\cos\ffrac{\pi s's}{p_-}\,\wkappap_{0,s'}(\tau)
    \\
    + 2\sum_{(r',s')\in\setii}\!\!\bigl(\cos\ffrac{\pi r's}{p_+}
    \cos\ffrac{\pi s's}{p_-}\,\wkappap_{r',s'}(\tau)
    - \sin\ffrac{\pi r's}{p_+}\sin\ffrac{\pi s's}{p_-}\,
    \wkappam_{r',s'}(\tau)\bigr)\!
    \biggr).
  \end{multline*}
  Taking the sum and the difference of these formulas with $s\mapsto
  p_-r-p_+s$ and $s\mapsto p_-r+p_+s$, we obtain~\eqref{eq:S-kappam}
  and~\eqref{eq:S-kappap}.
  
  Substituting
  \begin{equation*}
    g(r)=e^{-i\pi\frac{rs}{p_+p_-}},\quad
    h_+(r)=\ffrac{\tau}{\sqrt{\mathstrut 2p_+p_-}}
    \ffrac{\theta'_r(\tau)}{\eta(\tau)},
    \quad
    u^+_-(r,s)=\wrhop_{r,s},\quad
    u^-_-(r,s)=\wrhom_{r,s}
  \end{equation*}
  in~\eqref{eq:the-sum}, we find the transformation of the
  theta-constant derivatives as
  \begin{multline*}
    \frac{\theta'_{s}(-\frac{1}{\tau})}{\eta(-\frac{1}{\tau})}
    =\ffrac{-2i\tau}{\sqrt{\mathstrut 2p_+p_-\!}}
    \biggl(
    \sum_{(r',s')\in\setii}\!\!\!
    \bigl(\sin\ffrac{\pi r's}{p_+}\cos\ffrac{\pi s's}{p_-}
    \,\wrhop_{r',s'}(\tau)
    +\sin\ffrac{\pi s's}{p_-}\cos\ffrac{\pi r's}{p_+}
    \,\wrhom_{r',s'}(\tau)
    \bigr)\\
    +\half\sum_{r'=1}^{p_+-1}\sin\ffrac{\pi r's}{p_+}
    \,\wrhop_{r',0}(\tau)
    +\half\sum_{s'=1}^{p_--1}\sin\ffrac{\pi s's}{p_-}
    \,\wrhom_{0,s'}(\tau)\!
    \biggr).
  \end{multline*}  
  Taking the sum and the difference of these formulas with $s\mapsto
  p_-r-p_+s$ and $s\mapsto p_-r+p_+s$, we obtain~\eqref{eq:S-rhom},
  \eqref{eq:S-phim} and~\eqref{eq:S-rhop}, \eqref{eq:S-phip}.
  
  For the second derivatives of the theta-constants, we have
  \begin{multline*}
    \ffrac{1}{\eta(-\frac{1}{\tau})}
    \Bigl(\theta''_{p_+s+p_-r}(-\ffrac{1}{\tau})
    -\theta''_{p_+s-p_-r}(-\ffrac{1}{\tau})\Bigr)=\\
    \shoveleft{\quad{}
      =\ffrac{\sqrt{2}}{\sqrt{\mathstrut p_+p_-}}\ffrac{1}{\eta(\tau)}
      \sum_{(r',s')\in\setii}\!\!
      (-1)^{rs'+sr'+1}
      \sin\!\ffrac{\pi r's}{p_+}\,\sin\ffrac{\pi s's}{p_-}}
    \\
    {}\times\Bigl(
    \tau^2(\theta''_{p_+s'+p_-r'}(\tau)-\theta''_{p_+s'-p_-r'}(\tau))
    +i\pi p_+p_-\tau
    (\theta_{p_+s'+p_-r'}(\tau)-\theta_{p_+s'-p_-r'}(\tau))\Bigr),
  \end{multline*}
  whence \eqref{eq:S-xi}, \eqref{eq:S-psi}, and~\eqref{eq:S-zeta}
  immediately follow.
\end{proof}

\begin{lemma}
  \begin{align}
    \wkappam_{r,s}(\tau\!+\!1)={}&\lambda_{r,s}\wkappam_{r,s}(\tau),&
    \wkappap_{r,s}(\tau\!+\!1)={}&\lambda_{r,s}\wkappap_{r,s}(\tau),\\
    \wrhop_{r,s}(\tau\!+\!1)={}&\lambda_{r,s}\wrhop_{r,s}(\tau),&
    \wphip_{r,s}(\tau\!+\!1)={}&
    \lambda_{r,s}(\wphip_{r,s}(\tau)+\wrhop_{r,s}(\tau)),\\
    \wrhom_{r,s}(\tau\!+\!1)={}&\lambda_{r,s}\wrhom_{r,s}(\tau),&
    \wphim_{r,s}(\tau\!+\!1)={}&
    \lambda_{r,s}(\wphim_{r,s}(\tau)+\wrhom_{r,s}(\tau)),\\
    \wxi_{r,s}(\tau\!+\!1)={}&\lambda_{r,s}\wxi_{r,s}(\tau),&
    \wpsi_{r,s}(\tau\!+\!1)={}&
    \lambda_{r,s}(\wpsi_{r,s}(\tau)+2\wxi_{r,s}(\tau)),\\
    \wzeta_{r,s}(\tau\!+\!1)={}&
    \lambda_{r,s}(\wzeta_{r,s}(\tau)+\wpsi_{r,s}(\tau)
    +\wxi_{r,s}(\tau)),\kern-80pt&&
  \end{align}
  where
  \begin{equation}
    \lambda_{r,s}=e^{2i\pi(\Delta_{r,s}-\frac{c}{24})}
    =(-1)^{rs}e^{i\pi(\frac{p_-}{2p_+}r^2
      +\frac{p_+}{2p_-}s^2-\frac{1}{12})}.
  \end{equation}
\end{lemma}
\begin{proof}
  Elementary calculation.
\end{proof}

\subsubsection{}\label{sec:rho2chi} It may be useful to give the
formulas inverse to those in~\bref{sec:chi2rho}.  Let $\mathscr{I}$
denote the set of indices
\begin{equation*}
  \mathscr{I}= \{(r,s)\mid
  1\leq r\leq p_+\!-\!1,\quad 1\leq s\leq p_-\!-\!1\}
\end{equation*}
(actually labeling the Kac-table boxes) and let
$\bar{\mathscr{I}}_1=\mathscr{I}\backslash\setii$.\ \ For
$(r,s)\in\bar{\mathscr{I}}_1$, it is convenient to set
\begin{alignat*}{3}
  \wxi_{r,s}&=\wxi_{p_+-r,p_--s},&\quad
  \wrhom_{r,s}&=-\wrhom_{p_+-r,p_--s},&\quad
  \wrhop_{r,s}&=-\wrhop_{p_+-r,p_--s},\\
  \wkappap_{r,s}&= \wkappap_{p_+-r,p_--s},&\quad
  \wkappam_{r,s}&=\wkappam_{p_+-r,p_--s}.
\end{alignat*}
Then the formulas that invert those in~\bref{sec:chi2rho} are
\begin{align*}
  \chi^+_{r,s}={}&
  \ffrac{1}{(p_+p_-)^2}\Bigl(\wxi_{r,s}-p_-r
  \wrhom_{r,s}-p_+s\wrhop_{r,s} \\*
  &{}+\ffrac{p_+p_-rs}{2}\wkappap_{r,s}
  -\ffrac{p_+^2s^2+p_-^2r^2}{4}\wkappam_{r,s}\Bigr),\quad
  (r,s) \in\mathscr{I},\\
  \chi^+_{p_+,s}={}&
  \ffrac{1}{p_+p_-}\bigl(\wrhom_{0,p_--s}
  + \ffrac{p_+ s}{2}\,\wkappap_{0,p_--s}\bigr),
  \quad1\leq s\leq p_-\!-\!1,\\
  \chi^+_{r,p_-}={}&
  \ffrac{1}{p_+p_-}\bigl(\wrhop_{p_+-r,0}
  + \ffrac{p_- r}{2}\,\wkappap_{p_+-r,0}\bigr),
  \quad1\leq r\leq p_+\!-\!1, \\
  \chi^+_{p_+,p_-}={}&\half\,\wkappap_{0,0}, \\
  \chi^-_{r,s}={}& \ffrac{1}{(p_+p_-)^2}\Bigl(
  - \wxi_{p_+-r,s}-p_-r\wrhom_{p_+-r,s}+p_+s\wrhop_{p_+-r,s}\\*
  &{} +\ffrac{p_+p_-rs}{2}\wkappap_{p_+-r,s}
  +\ffrac{p_+^2s^2+p_-^2r^2-(p_+p_-)^2}{4}\wkappam_{p_+-r,s}
  \Bigr),\quad(r,s)\in\mathscr{I},\\
  \chi^-_{p_+,s}={}&
  \ffrac{1}{p_+p_-}\bigl(-\wrhom_{0,s}
  + \ffrac{p_+s}{2}\,\wkappap_{\,0,s}\bigr),
  \quad1\leq s\leq p_-\!-\!1,\\
  \chi^-_{r,p_-}={}&
  \ffrac{1}{p_+p_-}\bigl(-\wrhop_{r,0}
  + \ffrac{p_-r}{2}\,\wkappap_{r,0}\bigr),
  \quad1\leq r\leq p_+\!-\!1, \\
  \chi^-_{p_+,p_-}={}&\half\wkappap_{p_+,0}. 
\end{align*}

\subsection{Modular invariants}\label{sec:mod-inv}
The decomposition of the $\SLiiZ$ action established
in~\bref{thm:R-decomp} considerably simplifies finding sesquilinear
modular invariants.  We illustrate this by giving several easily
constructed series.

First, in the $\wrhop$, $\wrhom$, and $\wxi$ sectors (see
Table~\ref{table1}), modular invariants necessarily involve $\tau$
explicitly: they are given by
\begin{equation*}
  \boldsymbol{\wxi}^\boxslash(\tau,\bar\tau)
  =\sum_{r=1}^{p_+-1}\im\tau\,
  |\wrhop_{r,0}(\tau)|^2
  +2\sum_{(r,s)\in\setii}\im\tau\,
  |\wrhop_{r,s}(\tau)|^2,
\end{equation*}
a ``symmetric'' expression with the $\wrhom_{r,s}$, and
\begin{multline*}
  \boldsymbol{\wxi}(\tau,\bar\tau)
  =\sum_{(r,s)\in\setii}\bar\wxi_{r,s}( {\bar\tau }) ( 8(\im\tau)^2
  \wxi_{r,s}(\tau) + 4 p_+ p_-{\im\tau\, }\pi
  \chi_{r,s}(\tau) )\\*[-6pt]
  + \bar\chi_{r,s}( {\bar\tau }) ( 4p_+ p_-{\im\tau\, } \pi
  \wxi_{r,s}(\tau) + (\pi p_+ p_-)^2 \chi_{r,s}(\tau) ).
\end{multline*}
All these are expressed through the characters $\wkappam^\pm_{r,s}$
and $\chi_{r,s}$ in accordance with the formulas
in~\bref{sec:chi2rho}.

Next, in the $\wkappap$ sector, we have the $A$-series invariants
\begin{multline*}
  \boldsymbol{\wkappap}_{[A]}(\tau,\bar\tau) ={}\\
  {}=|\wkappap_{0,0}(\tau)|^2 +
  |\wkappap_{p_+,0}(\tau)|^2 + 2\sum_{r=1}^{p_+-1}
  |\wkappap_{r,0}(\tau)|^2 + 2\sum_{s=1}^{p_--1}
  |\wkappap_{0,s}(\tau)|^2 
  +4\!\sum_{(r,s)\in\setii}\!
  |\wkappap_{r,s}(\tau)|^2
\end{multline*}
and additional $D$-series invariants in the case where
$p_-\equiv0\,\mathrm{mod}\,4$:
\begin{multline*}
  \boldsymbol{\wkappap}_{[D]}(\tau,\bar\tau)
  =|\wkappap_{0, 0}(\tau) +
  \wkappap_{p_+, 0}(\tau)|^2 + \sum_{r=1}^{p_+ - 1} |\wkappap_{r,
    0}(\tau) + \wkappap_{p_+ - r, 0}(\tau)|^2
  \\*
  + \!\!\!\sum_{\substack{2\leq s\leq p_- - 1\\
      s\ \text{even}}}\!\!\!\!  |\wkappap_{0, s}(\tau) + \wkappap_{0,
    p_- - s}(\tau)|^2
  + \sum_{\substack{(r,s)\in\setii\\
      s\ \text{even}}}\!\!  2\,|\wkappap_{r, s}(\tau) + \wkappap_{r,
    p_- - s}(\tau)|^2.
\end{multline*}
Of course, ``symmetric'' $D$-invariants exist whenever
$p_+\equiv0\,\mathrm{mod}\,4$.  Again, all these invariants are
expressed through the characters via the formulas
in~\bref{sec:chi2rho}.

We also note an $E_6$-like invariant for $(p_+,p_-)=(5,12)$:
\begin{multline*}
  \boldsymbol{\wkappap}_{[E_6]}(\tau,\bar\tau) =|\wkappap_{0,1}(\tau)
  - \wkappap_{0,7}(\tau)|^2 + |\wkappap_{0,2}(\tau) -
  \wkappap_{0,10}(\tau)|^2 + |\wkappap_{0,5}(\tau) -
  \wkappap_{0,11}(\tau)|^2
  \\
  \shoveright{{}+ 2 |\wkappap_{1,1}(\tau) - \wkappap_{1,7}(\tau)|^2 +
    2 |\wkappap_{2,1}(\tau) - \wkappap_{2,7}(\tau)|^2 + 2
    |\wkappap_{2,5}(\tau) - \wkappap_{3,1}(\tau)|^2\ }
  \\
  + 2 |\wkappap_{2,2}(\tau) - \wkappap_{3,2}(\tau)|^2 + 2
  |\wkappap_{1,5}(\tau) - \wkappap_{4,1}(\tau)|^2 + 2
  |\wkappap_{1,2}(\tau) - \wkappap_{4,2}(\tau)|^2.
\end{multline*}

It would be quite interesting to systematically obtain all modular
invariants as quantum-group invariants.

\section{Conclusions}
In the logarithmically extended $(p_+,p_-)$ minimal models, we have
constructed the chiral vertex-operator algebra~$\talgW$, its
irreducible representations and Verma modules, and calculated the
$\SLiiZ$-representation generated by the characters.  The space of
generalized characters carrying that representation most probably
coincides with the space of torus amplitudes.  A great deal of work
remains to be done, however.

\subsubsection*{\textbf{\thesection.1}\ }
\begin{figure}[tbph]
  \includegraphics[bb=1.7in 8.1in 8.8in 10.3in, clip]{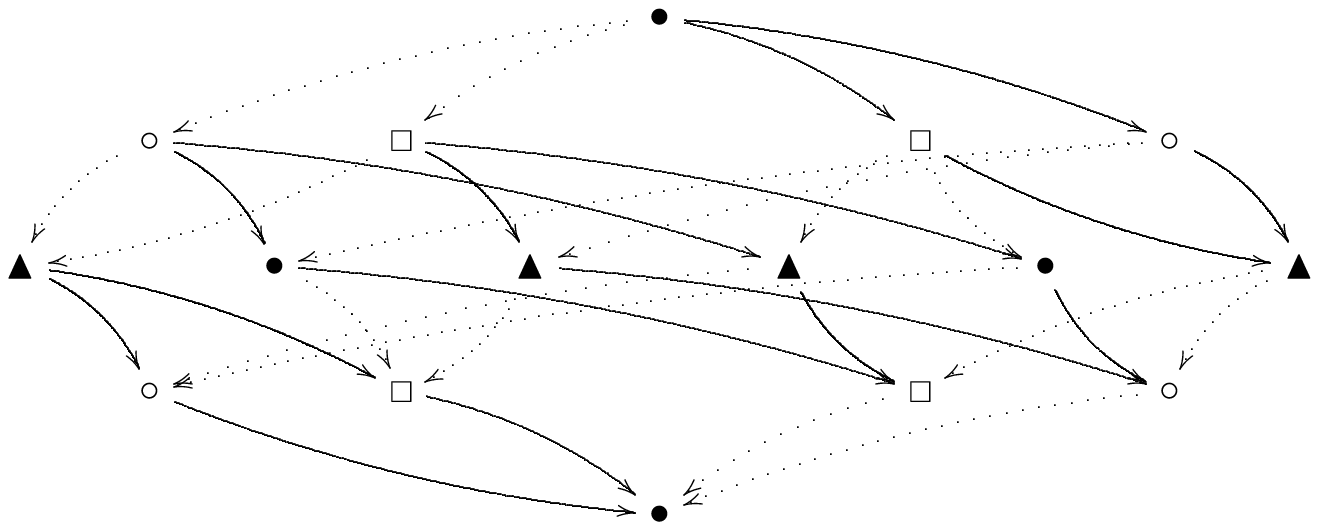}
  \caption[Sixteen subquotients]{\captionfont{Sixteen subquotients,
      which do not suffice to build a projective $\talgW$-module.}}
  \label{fig:qgprojective}
\end{figure}

First and foremost, constructing the chiral sector of the space of
states requires building projective modules of $\talgW$.
Unfortunately, little is known about their structure (some indirect
but quite useful information has recently become available
in~\cite{[Flohrpp]}).  One of the clues is the known
structure~\cite{[FGST-q]} of projective modules of the
Kazhdan--Lusztig-dual quantum group~$\XXX$, which must be a piece of
the structure of projective $\WWW$-modules.  That is, taking an
irreducible $\WWW$-module~$\repX^+_{r,s}$ (with $1\leq r\leq
p_+\!-\!1$ and $1\leq s\leq p_-\!-\!1$), replacing it with the
$\XXX$-module~$\XX^+_{r,s}$, taking the universal projective cover
$\mathsf{P}^+_{r,s}$ of the latter, and translating the result back
into the $\WWW$-language, we obtain the structure of sixteen
subquotients in Fig.~\ref{fig:qgprojective},
where solid and dotted lines denote the elements of $\mathrm{Ext}^1$
and the symbolic notation for the modules is as in Fig.~\ref{fig:Qrs},
i.e., $\bullet=\repX^{+}_{r,s}$, $\scrBox=\repX^{-}_{p_+-r,s}$,
$\circ=\repX^{-}_{r,p_--s}$, and
$\blacktriangle=\repX^{+}_{p_+-r,p_--s}$.  However, \textit{in
  addition to these subquotients and embeddings}, the projective
$\WWW$-module $\repP^+_{r,s}$ (the universal projective cover of
$\repX^+_{r,s}$) \textit{must also contain subquotients isomorphic to
  the minimal-model representations~$\repX_{r,s}$}.  Furthermore, the
$\WWW$-representation category contains the projective covers
$\repP_{r,s}$ of the irreducible representations $\repX_{r,s}$, which
have no projective-module counterparts in the $\XXX$-representation
category (the $\repP_{r,s}$ may be interesting because of their
relation to the boundary-condition-changing operator whose $4$-point
correlation function gives the Cardy formula~\cite{[Cardy]} for the
crossing probability).

\subsubsection*{\textbf{\thesection.2}\ }
The foregoing is directly relevant to establishing the relation
between our construction, specialized to $(p_+\!=\!3,p_-\!=\!2)$, and
the analysis in~\cite{[qCardy],[Cardy-qu],[GL-2002],[GL]
}, where a family of $c=0$ models was considered, labeled by a
parameter~$b$.  The approach in~\cite{[GL]}, in principle, allows
adding various primary fields to the theory and, in this sense, is not
aimed at fixing a particular chiral algebra.  The logarithmic minimal
models in this paper, on the other hand, are \textit{minimal
  extensions} of the rational $(p_+,p_-)$ models, minimal in the sense
that consistent ``intermediate'' models\,---\,with a subalgebra of
$\talgW$ but with a finite number of fields\,---\,do not seem to
exist.\footnote{To avoid misunderstanding, we note that logarithmic
  models may of course be constructed based on different $W$-algebras,
  overlapping with $\talgW$ only over the Virasoro algebra} There are
various ways to construct ``larger'' logarithmic extensions of minimal
models, for example, by taking the kernel not of two but of
\textit{one} screening (cf.~\cite{[FFHST]}).  In our ``minimal''
setting, in particular, the $\algW_{3,2}$-primary fields are just
those whose dimensions are in Table~\ref{table-kac}, and similarly for
all the $(p_+,p_-)$ models, as stated in~\bref{thm:R-decomp}.  (We
note once again that the \textit{Virasoro} field content, which
appears to have been the subject of some discussion in other
approaches, then follows uniquely from the decompositions
in~\bref{prop:x-vir}\,---\,in fact, from the structure of the relevant
complexes.  In particular, the number of Virasoro primary fields is
not limited to an integer multiple of either the standard
$(p_+\!-\!1)\times(p_-\!-\!1)$ or the extended $p_+\times p_-$ Kac
table.)  \ In the setting in this paper, fixing the value of~$b$ or
another similar parameter requires constructing projective
$\algW$-modules, which will be addressed elsewhere (among other
things, it will completely settle the ``logarithmic partner'' issues).

\subsubsection*{\textbf{\thesection.3}\ }
The $\SLiiZ$-representation on the generalized characters (presumably,
torus amplitudes) coincides with the $\SLiiZ$-representation on the
center of the Kazhdan--Lusztig-dual quantum
group~$\XXX$~\cite{[FGST-q]}.  This remarkable correspondence deserves
further study, as do other aspects of the Kazhdan--Lusztig
correspondence.  

The Kazhdan--Lusztig correspondence suggests that in some generality,
the space of torus amplitudes $\TA$ and the conformal field theory
center $\cZcft$ are related by conformal-field-theory analogues of the
Radford and Drinfeld maps known in the theory of quantum groups.
Then, under the identification of~$\cZcft$ with the space of boundary
conditions preserved by~$\algW$, the images of irreducible characters
under the ``$\algW$-Radford map'' are the Ishibashi states and the
images under the ``$\algW$-Drinfeld map'' are the Cardy states.

We also give a fusion algebra suggested by the Kazhdan--Lusztig
correspondence.  This fusion is the $\XXX$ Grothendieck ring
$\Grring$, with the preferred basis of $2p_+ p_-$ irreducible
representations.  In terms of the $\WWW$ algebra, we identify the
preferred basis elements with the $2p_+ p_-$ representations
$\repK^\pm_{r,s}$ defined in~\eqref{eq:ker-scr}.  For all $1\leq
r,r'\leq p_+$, $1\leq s,s'\leq p_-$, and $\alpha,\beta=\pm$, we thus
have the algebra
\begin{equation}\label{the-fusion}
  \repK^{\alpha}_{r,s}\repK^{\beta}_{r',s'}
  =\sum_{\substack{u=|r - r'| + 1\\
      \mathrm{step}=2}}^{r + r' - 1}
  \sum_{\substack{v=|s - s'| + 1\\
      \mathrm{step}=2}}^{s + s' - 1}
  {\tilde{\repK}}^{\alpha\beta}_{u,v},
\end{equation}
where
\begin{equation*}
  {\tilde{\repK}}^{\alpha}_{r,s} =
  \begin{cases}
    \repK^{\alpha}_{r,s},& 
    \begin{array}[t]{l}
      1\leq r\leq p_+,\\[-6pt]
      1\leq s\leq p_-,
    \end{array}
    \\
    \repK^{\alpha}_{2p_+ - r,s} + 2\repK^{-\alpha}_{r - p_+, s},&
    \begin{array}[t]{l}
      p_+\!+\!1\leq r\leq 2 p_+\!-\!1,\\[-6pt]
      1\leq s\leq p_-,
    \end{array}
    \\
    \repK^{\alpha}_{r,2p_- - s} + 2\repK^{-\alpha}_{r,s - p_-},&
    \begin{array}[t]{l}
      1\leq r\leq p_+,\\[-6pt]
      p_-\!+\!1\leq s\leq 2 p_-\!-\!1,
    \end{array}
    \\
    \mbox{}\kern-3pt\begin{aligned}[b]
      &\repK^{\alpha}_{2p_+ - r, 2 p_- - s}
      + 2\repK^{-\alpha}_{2p_+ - r, s - p_-}\\[-3pt]
      &{}+ 2\repK^{-\alpha}_{r - p_+, 2 p_- - s}
      + 4\repK^{\alpha}_{r - p_+, s - p_-},
    \end{aligned}&
    \begin{array}[t]{l}
      p_+\!+\!1\leq r\leq 2 p_+\!-\!1,\\[-6pt]
      p_-\!+\!1\leq s\leq 2 p_-\!-\!1.
    \end{array}
  \end{cases}
\end{equation*}
This algebra has several noteworthy properties:
\begin{enumerate}
\item it is generated by two elements $\repK^+_{1,2}$ and
  $\repK^+_{2,1}$;
  
\item its radical is generated by the algebra action on
  $\repK^+_{p_+,p_-}$; the quotient over the radical coincides with
  the fusion of the $(p_+,p_-)$ Virasoro minimal models;
  
\item  $\repK^+_{1,1}$ is the identity;
  
\item $\repK^-_{1,1}$ acts as a simple current,
  $\repK^-_{1,1}\repK^\alpha_{r,s}=\repK^{-\alpha}_{r,s}$.  
\end{enumerate}

A functor between the representation categories of $\WWW$ and $\XXX$
is not yet known, and the identification of the basis of irreducible
$\XXX$-representation with the $\repK^\pm_{r,s}$ is based on indirect
arguments.  First, it is clear that the minimal-model representations
$\repX_{r,s}$ act by zero on the other basis elements in the fusion
algebra, simply because of the vanishing of three-point functions
involving two minimal-model representations and one representation on
which the ideal $\algR$ acts nontrivially.  We next recall that the
$2p_+ p_-$ representations $\repK^\pm_{r,s}$ are defined as kernels of
the screenings.  Some of them are reducible, see~\eqref{K-reducile},
``inasmuch as'' the Felder complexes have a nonzero cohomology (the
minimal-model representations $\repJ_{r,s}\equiv\repX_{r,s}$).  Their
role as counterparts of the irreducible $\XXX$-representa\-tions under
the Kazhdan--Lusztig correspondence is supported by the numerical
evidence in~\cite{[Flohrpp]}: decomposing $\repK^\pm_{r,s}$ into the
Virasoro modules (see~\bref{lemma:sl-vir}) shows that the $(3,2)$ and
$(5,2)$ results in~\cite{[Flohrpp]} agree with~\eqref{the-fusion}.
Needless to say, it would be quite interesting to properly define the
$\WWW$-fusion and establish~\eqref{the-fusion} by some ``relatively
direct'' (or just numerical, as the first step) calculation of the
$\WWW$ (not just Virasoro) coinvariants.\footnote{The fusion algebra
  in~\eqref{the-fusion} also follows from the above $\SLiiZ$ action on
  the characters via a procedure generalizing the Verlinde formula,
  similar to that in~\cite{[FHST]}; the (somewhat bulky) details will
  be given elsewhere.}

We also note that the quantum group $\XXX$ can be quite useful in
constructing the full (holomorphic${}+{}$antiholomorphic) space of
states, by taking $\XXX$-invariants in the product of
$(\talgW,\tqalgA)$-bimodules.

\subsubsection*{Acknowledgments} 
We are grateful to A.~Belavin, J.~Fuchs, A.~Isaev, S.~Parkhomenko, and
P.~Pyatov for the useful discussions and comments.  This paper was
supported in part by the RFBR Grant 04-01-00303 and the RFBR--JSPS
Grant 05-01-02934YaF\_a.  The work of AMG, AMS and IYuT is supported
in part by the LSS-4401.2006.2 grant.  The work of IYuT was supported
in part by the RFBR Grant 05-02-17217 and the ``Dynasty'' foundation.

\appendix

\section{$(3,2)$-model examples}\label{app:32-example}
In the simplest case of the $(3,2)$ model, the minimal-model character
is trivial, $\chi_{1,1}(q)=1$, and $12$ nontrivial characters are
expressed through theta-constants in accordance with
Eqs.~\eqref{prop-chi+_rs}--\eqref{prop-chi-_rs}.  Explicitly, for
example, the characters of $\repX^{\pm}_{1,1}$ are
\begin{align*}
  q^{-2}\chi^+_{1,1}(q)={}&1 + q + 2 q^2 + 2 q^3 + 4 q^4 + 4 q^5 + 
  7 q^6 + 8 q^7 + 12 q^8 + 14 q^9\\
   &\qquad{}+ 21 q^{10}  
   + 24 q^{11} + 34 q^{12} + 44 q^{13} + 58 q^{14} + 72 q^{15}
  +\dots\\
  \intertext{(with the character of $\repK_{1,1}^+$ given by
    $1+\chi^+_{1,1}(q)$), and}
  q^{-7}\chi^-_{1,1}(q)={}&2 + 2 q + 4 q^2 + 6 q^3 + 10 q^4 + 12 q^5 +
  20 q^6 + 26 q^7 + 36 q^8 + 48 q^9\\
  &\qquad{}+ 66 q^{10}  
  + 84 q^{11} + 114 q^{12} + 144 q^{13} + 188 q^{14} + 240 q^{15}
  + \dots
\end{align*}
(the dimensions of the respective highest-weight vectors are
$\Delta_{1,1;1}=2$ and $\Delta_{2,1;-2}=7$).  The modular
transformation properties of the characters follow by expressing them
through the basis in Table~\ref{table1} (where now $|\setii|=1$,
$|\seti|=6$, $|\setm|=2$, and $|\setp|=3$) in accordance with the
formulas in~\bref{sec:rho2chi} and using~\bref{S-lemma}.  Under the
$\SLiiZ$ action, the $13$ characters (including $\chi_{1,1}$) give
rise to the dimension-$20$ space of generalized characters.  To give
examples of modular transformations, we use Eqs.~\eqref{varphi-def}
and express the $\modS$-transformed characters through the characters
with $\tau$-dependent coefficients:
\begin{multline*}
  \chi^+_{1,1}(-\ffrac{1}{\tau})
  =
  \bigl(-\ffrac{13}{144} 
  + \ffrac{i({\sqrt{3}} + 18\pi)\tau}{108}  - 
  \ffrac{{\tau }^2}{144} \bigr)\chi_{1,1}(\tau)
  + \bigl(\ffrac{i\tau }{6{\sqrt{3}}}
  - \ffrac{{\tau }^2}{3} \bigr) \chi^-_{1,1}(\tau)\\
  {}+ \bigl(\ffrac{1}{12{\sqrt{3}}} + \ffrac{i\tau}{6}\bigr)
  \chi^-_{1,2}(\tau)
  - \bigl(\ffrac{i\tau }{6{\sqrt{3}}} + \ffrac{{\tau }^2}{6} \bigr)
  \chi^-_{2,1}(\tau)
  + \bigl(\ffrac{i\tau}{12} - \ffrac{1}{12{\sqrt{3}}}\bigr)
  \chi^-_{2,2}(\tau)
  - \ffrac{i\tau}{6{\sqrt{3}}}\,\chi^-_{3,1}(\tau)\\
  {}- \ffrac{1} {12{\sqrt{3}}}\,\chi^-_{3,2}(\tau)
  + \bigl(\ffrac{{\tau}^2}{3} - \ffrac{i\tau}{6{\sqrt{3}}}\bigr)
  \chi^+_{1,1}(\tau)
  - \bigl(\ffrac{1}{12{\sqrt{3}}} + \ffrac{i\tau}{6}\bigr)
  \chi^+_{1,2}(\tau)
  + \bigl(\ffrac{i\tau } {6{\sqrt{3}}} + \ffrac{{\tau }^2}{6}\bigr) 
  \chi^+_{2,1}(\tau)\\
  {}+ \bigl(\ffrac{1}{12{\sqrt{3}}}
  - \ffrac{i\tau}{12}\bigr) \chi^+_{2,2}(\tau)
  + \ffrac{i\tau}{6{\sqrt{3}}}\,\chi^+_{3,1}(\tau)
  + \ffrac{1}{12{\sqrt{3}}}\,\chi^+_{3,2}(\tau)
\end{multline*}
and
\begin{multline*}
  \chi^-_{1,1}(-\ffrac{1}{\tau})
  =
  \bigl(-\ffrac{23}{144} - \ffrac{i( {\sqrt{3}} + 18\pi)\tau}{108}   + 
  \ffrac{{\tau }^2}{144} \bigr) \chi_{1,1}(\tau)
  + \bigl(\ffrac{{\tau }^2}{3} - \ffrac{i\tau }{6{\sqrt{3}}}\bigr)
  \chi^-_{1,1}(\tau)\\
  {}+ \bigl( \ffrac{1}{12{\sqrt{3}}} + 
  \ffrac{i\tau }{6}\bigr) \chi^-_{1,2}(\tau) + 
  \bigl(\ffrac{i\tau } {6{\sqrt{3}}} + 
  \ffrac{{\tau }^2}{6}\bigr) \chi^-_{2,1}(\tau)
  + \bigl(\ffrac{i \tau}{12} -  \ffrac{1}{12{\sqrt{3}}}\bigr)
  \chi^-_{2,2}(\tau)
  + \ffrac{i\tau }{6{\sqrt{3}}}\,\chi^-_{3,1}(\tau)\\
  {}- \ffrac{1} {12{\sqrt{3}}}\,\chi^-_{3,2}(\tau)
  + \bigl( \ffrac{i\tau } {6{\sqrt{3}}} - 
  \ffrac{{\tau }^2}{3}\bigr) \chi^+_{1,1}(\tau)
  - \bigl(\ffrac{1}{12{\sqrt{3}}} + \ffrac{i\tau}{6}\bigr)
  \chi^+_{1,2}(\tau)
  - \bigl( \ffrac{i\tau }{6{\sqrt{3}}} +
  \ffrac{{\tau }^2}{6} \bigr) \chi^+_{2,1}(\tau)\\
  {}+ \bigl( \ffrac{1}{12{\sqrt{3}}} - 
  \ffrac{i \tau}{12}\bigr) \chi^+_{2,2}(\tau) - 
  \ffrac{i\tau }{6{\sqrt{3}}}\,\chi^+_{3,1}(\tau)
  + \ffrac{1} {12{\sqrt{3}}}\,\chi^+_{3,2}(\tau).
\end{multline*}

We next consider fusion relations~\eqref{the-fusion}.  To write them
explicitly for $(p_+,p_-)=(3,2)$, we recall that $\repK^-_{1,1}$ acts
as $\repK^\alpha_{r,s}\repK^-_{1,1}=\repK^{-\alpha}_{r,s}$, and
therefore the entire $12\times 12$ multiplication table essentially
reduces to its $6\times 6$ block, where (recalling that
$\repK^+_{1,1}$ acts as identity) the $\half\cdot 5\cdot6$ independent
relations are
\begin{alignat*}{3}
  \repK^+_{1, 2} \repK^+_{1, 2}&= 
  2 \repK^-_{1, 1} + 2 \repK^+_{1, 1},\quad&
  \repK^+_{1, 2}\repK^+_{2, 1}&= \repK^+_{2, 2},\quad&
  \repK^+_{1, 2} \repK^+_{2, 2}&= 2 \repK^-_{2, 1} + 2 \repK^+_{2, 1},\\
  \repK^+_{1, 2} \repK^+_{3, 1}&= \repK^+_{3, 2},\quad&
  \repK^+_{1, 2} \repK^+_{3, 2}&= 2 \repK^-_{3, 1} + 2 \repK^+_{3, 1},
  \quad&
  \\
  \repK^+_{2, 1} \repK^+_{2, 1}&= \repK^+_{1, 1} + \repK^+_{3, 1},\quad&
  \repK^+_{2, 1} \repK^+_{2, 2}&= \repK^+_{1, 2} + \repK^+_{3, 2},\quad&
  \repK^+_{2, 1} \repK^+_{3, 1}&= 2 \repK^-_{1, 1} + 2 \repK^+_{2, 1},\\
  \repK^+_{2, 1} \repK^+_{3, 2}&= 2 \repK^-_{1, 2} + 2 \repK^+_{2, 2},
  \quad&
\end{alignat*}
\vspace{-1.3\baselineskip}
\begin{alignat*}{2}
  \repK^+_{2, 2} \repK^+_{2, 2}&= 2 \repK^-_{1, 1} + 2 \repK^-_{3, 1}
  + 2 \repK^+_{1, 1} + 2 \repK^+_{3, 1},\qquad&
  \repK^+_{2, 2} \repK^+_{3, 1}&= 2 \repK^-_{1, 2} + 2 \repK^+_{2, 2},\\
  \repK^+_{2, 2} \repK^+_{3, 2}&= 4 \repK^-_{1, 1} + 4 \repK^-_{2, 1}
  + 4 \repK^+_{1, 1} + 4 \repK^+_{2, 1},\qquad&
  \\
  \repK^+_{3, 1} \repK^+_{3, 1}&= 2 \repK^-_{2, 1} + 2 \repK^+_{1, 1}
  + \repK^+_{3, 1},\qquad&
  \repK^+_{3, 1} \repK^+_{3, 2}&= 2 \repK^-_{2, 2} + 2 \repK^+_{1, 2}
  + \repK^+_{3, 2},\kern16pt\\
  \repK^+_{3, 2} \repK^+_{3, 2}&= 4 \repK^-_{1, 1} + 4 \repK^-_{2, 1} + 
  2 \repK^-_{3, 1} + 4 \repK^+_{1, 1} + 4 \repK^+_{2, 1} + 
  2 \repK^+_{3, 1}.\kern-100pt
\end{alignat*}

\begin{small}
  {\normalsize Finally, more as a curiosity than for any practical
    purposes, we give explicit free-field expressions for the
    $\algW_{3,2}$-algebra generators in~\eqref{W-pm} (arbitrarily
    normalized):}
\begin{multline*}
  W^+=\Bigl(\sfrac{35}{27}\bigl(\dd^{4}\varphi\bigr)^2
  + \sfrac{56}{27}\,\dd^{5}\varphi\,\dd^{3}\varphi
  + \sfrac{28}{27}\,\dd^{6}\varphi\,\dd^{2}\varphi
  + \sfrac{8}{27}\,\dd^{7}\varphi\,\dd\varphi
  - \sfrac{280}{9 {\sqrt{3}}}\bigl(\dd^{3}\varphi\bigr)^2
  \,\dd^{2}\varphi
  \\
  {}- \sfrac{70}{3 {\sqrt{3}}}\,\dd^{4}\varphi\,
  \bigl(\dd^{2}\varphi\bigr)^2
  - \sfrac{280}{9 {\sqrt{3}}}\,\dd^{4}\varphi\,\dd^{3}\varphi
  \,\dd\varphi
  - \sfrac{56}{3 {\sqrt{3}}}\,\dd^{5}\varphi\,\dd^{2}\varphi
  \,\dd\varphi
  - \sfrac{28}{9 {\sqrt{3}}}\,\dd^{6}\varphi\bigl(\dd\varphi\bigr)^2
  \\
  {}+ 
  \sfrac{35}{3}\bigl(\dd^{2}\varphi)^4
  + \sfrac{280}{3}\,\dd^{3}\varphi\bigl(\dd^{2}\varphi\bigr)^2\,
  \dd\varphi + 
  \sfrac{280}{9}\bigl(\dd^{3}\varphi)^2
  \bigl(\dd\varphi\bigr)^2
  + \sfrac{140}{3}\,\dd^{4}\varphi\,\dd^{2}\varphi
  (\dd\varphi\bigr)^2
  \\
  {}+ \sfrac{56}{9}\,\dd^{5}\varphi\bigl(\dd\varphi\bigr)^3
  - \sfrac{140}{{\sqrt{3}}}\bigl(\dd^{2}\varphi\bigr)^3
  \bigl(\dd\varphi\bigr)^2
  - \sfrac{560}{3 {\sqrt{3}}}\,\dd^{3}\varphi\,\dd^{2}\varphi \,
  \bigl(\dd\varphi\bigr)^2
  - \sfrac{70}{3 {\sqrt{3}}}\,\dd^{4}\varphi
  \bigl(\dd\varphi\bigr)^4
  \\
  {}+ 70 \bigl(\dd^{2}\varphi\bigr)^2
  \bigl(\dd\varphi\bigr)^4
  + \sfrac{56}{3}\,\dd^{3}\varphi
  \bigl(\dd\varphi)^5
  - \sfrac{28}{{\sqrt{3}}}\,\dd^{2}\varphi
  \bigl(\dd\varphi\bigr)^6
  + \bigl(\dd\varphi)^8
  - \sfrac{1}{27 {\sqrt{3}}}\,\dd^{8}\varphi\Bigr)e^{2\sqrt{3}\varphi},
\end{multline*}
\vspace{-.9\baselineskip}
\begin{multline*}
  W^-=
  \Bigl(
  \sfrac{217}{192} \bigl(\dd^{5}\varphi\bigr)^2
  - \sfrac{2653}{3456}\,\dd^{6}\varphi \,\dd^{4}\varphi
  - \sfrac{23}{384}\,\dd^{7}\varphi \,\dd^{3}\varphi
  - \sfrac{11}{1152} \,\dd^{8}\varphi \,\dd^{2}\varphi
  - \sfrac{1}{768}\,\dd^{9}\varphi \,\dd\varphi \\
  {}- 
  \sfrac{1225}{64 {\sqrt{3}}}\,\dd^{4}\varphi
  \bigl(\dd^{3}\varphi\bigr)^2
  - \sfrac{13475}{576 {\sqrt{3}}}\bigl(\dd^{4}\varphi\bigr)^2
  \,\dd^{2}\varphi
  + \sfrac{2695}{64 {\sqrt{3}}}\,\dd^{5}\varphi \,\dd^{3}\varphi
  \,\dd^{2}\varphi +  
  \sfrac{2555}{192 {\sqrt{3}}}\,\dd^{5}\varphi \,\dd^{4}\varphi
  \,\dd\varphi \\
  {}- 
  \sfrac{2891}{576 {\sqrt{3}}}\,\dd^{6}\varphi
  \bigl(\dd^{2}\varphi\bigr)^2
  - \sfrac{1351}{192 {\sqrt{3}}}\,\dd^{6}\varphi \,\dd^{3}\varphi
  \,\dd\varphi -  
  \sfrac{103}{192 {\sqrt{3}}}\,\dd^{7}\varphi \,\dd^{2}\varphi
  \,\dd\varphi -  
  \sfrac{13}{384 {\sqrt{3}}}\,\dd^{8}\varphi
  \bigl(\dd\varphi\bigr)^2\\
  {}+ 
  \sfrac{3535}{32}\bigl(\dd^{3}\varphi\bigl)^2
  \bigl(\dd^{2}\varphi\bigr)^2
  - \sfrac{735}{16}\bigl(\dd^{3}\varphi\bigr)^3\,\dd\varphi
  - \sfrac{3395}{54}\,\dd^{4}\varphi
  \bigr(\dd^{2}\varphi\bigl)^3
  + \sfrac{245}{24}\,\dd^{4}\varphi \,\dd^{3}\varphi \,\dd^{2}\varphi
  \,\dd\varphi\\
  {}+ \sfrac{12635}{576}\bigl(\dd^{4}\varphi\bigr)^2
  \bigl(\dd\varphi\bigr)^2
  + \sfrac{245}{12}\,\dd^{5}\varphi \bigl(\dd^{2}\varphi\bigr)^2
  \,\dd\varphi
  + \sfrac{105}{32}\,\dd^{5}\varphi \,\dd^{3}\varphi
  \bigl(\dd\varphi\bigr)^2\\
  {}- \sfrac{2443}{288}\,\dd^{6}\varphi \,\dd^{2}\varphi
  \bigl(\dd\varphi\bigr)^2
  - \sfrac{19}{96}\,\dd^{7}\varphi \bigl(\dd\varphi\bigr)^3
  - \sfrac{13405}{144 {\sqrt{3}}}\bigl(\dd^{2}\varphi\bigr)^5
  + \sfrac{8225}{24 {\sqrt{3}}}\,\dd^{3}\varphi
  \bigl(\dd^{2}\varphi)^3 \,\dd\varphi\\
  {}- \sfrac{105 {\sqrt{3}}}{4}\bigl(\dd^{3}\varphi\bigr)^2
  \,\dd^{2}\varphi \bigl(\dd\varphi\bigr)^2
  + \sfrac{665}{24 {\sqrt{3}}}\,\dd^{4}\varphi
  \bigl(\dd^{2}\varphi\bigr)^2 \bigl(\dd\varphi\bigr)^2
  + \sfrac{245}{2 {\sqrt{3}}}\,\dd^{4}\varphi \,\dd^{3}\varphi
  \bigl(\dd\varphi\bigr)^3
  \\
  {}- \sfrac{245}{8 {\sqrt{3}}}\,\dd^{5}\varphi \,\dd^{2}\varphi
  \bigl(\dd\varphi\bigr)^3
  - \sfrac{91}{24 {\sqrt{3}}}\,\dd^{6}\varphi
  \bigl(\dd\varphi\bigr)^4
  + \sfrac{16205}{144}
  \bigl(\dd^{2}\varphi\bigr)^4
  \bigl(\dd\varphi\bigr)^2
  + \sfrac{385}{4}\,\dd^{3}\varphi \bigl(\dd^{2}\varphi\bigr)^2
  \bigl(\dd\varphi\bigr)^3
  \\
  {}+ \sfrac{525}{8}\bigl(\dd^{3}\varphi\bigr)^2
  \bigl(\dd\varphi\bigr)^4
  + \sfrac{35}{3}\,\dd^{4}\varphi \,\dd^{2}\varphi
  \bigl(\dd\varphi\bigr)^4
  - 7 \,\dd^{5}\varphi
  \bigl(\dd\varphi\bigr)^5
  + \sfrac{665}{3 {\sqrt{3}}}\bigl(\dd^{2}\varphi\bigr)^3
  \bigl(\dd\varphi\bigr)^4\\
  {}+ \sfrac{105 {\sqrt{3}}}{2}\,\dd^{3}\varphi \,\dd^{2}\varphi
  \bigl(\dd\varphi\bigr)^5
  - \sfrac{35}{3 {\sqrt{3}}}\,\dd^{4}\varphi
  \bigl(\dd\varphi\bigr)^6
  + \sfrac{455}{6}\bigl(\dd^{2}\varphi\bigr)^2
  \bigl(\dd\varphi\bigr)^6
  + 5 \,\dd^{3}\varphi \bigl(\dd\varphi\bigr)^7
  \\
  {}+ \sfrac{25}{{\sqrt{3}}}\,\dd^{2}\varphi
  \bigl(\dd\varphi\bigr)^8
  + \bigl(\dd\varphi\bigr)^{10}
  - \sfrac{1}{13824 {\sqrt{3}}}\,\dd^{10}\varphi
  \Bigr)e^{-2\sqrt{3}\varphi},
\end{multline*}
{\normalsize where, despite the brackets introduced for the
  compactness of notation, the nested normal ordering is from right to
  left, e.g., $\,\dd^{4}\varphi(\dd^{2}\varphi(\dd^{2}\varphi
  e^{2\sqrt{3}\varphi}))$.}
\end{small}
These dimension-$15$ operators have the OPE
\begin{equation*}
  W^+(z)\,W^-(w)=2^7\cdot3\cdot5^3\cdot7^2\cdot11\cdot17\,
  \ffrac{T(w)}{(z-w)^{28}} + \dots,
\end{equation*}
with the energy--momentum tensor given by~\eqref{eq:the-Virasoro},
that is,
\begin{equation*}
   T(z)=\half\,\dd\varphi(z)\dd\varphi(z)
   - \ffrac{1}{2\sqrt{3}}\,\dd^2\varphi(z).
\end{equation*}
In the $(3,2)$ model, the minimal-model vertex-operator algebra
$\talgM$ is trivial, or in other words, $T(z)$ is in the
ideal~$\algR$.

\section{Quantum-group technicalities}\label{app:QG}

\subsection{Drinfeld double of a quantum group}\label{app:double}
We recall~\cite{[Kassel],[ChP]} that the space $H^*$ of linear
functions on a Hopf algebra $H$ is a Hopf algebra with the
multiplication, comultiplication, unit, counit, and antipode given by
\begin{equation}\label{double-def}
  \begin{gathered}
    \coup{\beta\gamma}{x}=\sum_{(x)}\coup{\beta}{x'}
    \coup{\gamma}{x''},\quad
    \coup{\Delta(\beta)}{x\tensor y}=\coup{\beta}{yx},\\
    \coup{\one}{x}=\epsilon(x),\quad
    \epsilon(\beta)=\coup{\beta}{\one},\quad
    \coup{S(\beta)}{x}=\coup{\beta}{S^{-1}(x)}
  \end{gathered}
\end{equation}
for any $\beta,\gamma\,{\in}\, H^*$ and $x,y\,{\in}\, H$.
The Drinfeld double~\cite{{[Kassel]},[ChP]} $D(H)$ is a Hopf
algebra with the underlying vector space $H^*\tensor H$ and
with the multiplication, comultiplication, unit, counit, and antipode
given (in addition to the formulas for $H$ and $H^*$) by
\begin{equation}\label{double-def-1}
  x\beta=\sum_{(x)}\beta(S^{-1}(x''')?x')x'',
  \qquad x\in H,\quad\beta\in H^*.
\end{equation}

\subsection{Proof of~\bref{thm:double}}\label{app:double-proof}
By induction, it is easy to see that the comultiplication in the PBW
basis in~$\qalgB$ is given by
\begin{multline}\label{multiplic}
  \Delta( \pbw_{jmn})={}\\
  =\sum_{r=0}^m\sum_{s=0}^n{\qbin{m}{r}}_+{\qbin{n}{s}}_- q_+^{p_- r(r
    - m)} q_-^{p_+ s(s - n)}
  \pbw_{j+2p_-(m-r)-2p_+(n-s),r,s}\tensor\pbw_{j,m-r,n-s}.
\end{multline}

With~\eqref{ejmn}, we define $\dkk, \fp, \emi \in\qalgB^*$ by the
relations
\begin{equation}\label{basis}
  \begin{gathered}
    \coup{\dkk}{\pbw_{jmn}}=\delta_{m,0}\delta_{n,0}\q^{j},\\
    \coup{\fp}{\pbw_{jmn}}
    =-\delta_{m,1}\delta_{n,0}\ffrac{\qp^{j}}{\qp^{p_-}\!-
      \qp^{-p_-}}, \quad \coup{\emi}{\pbw_{jmn}}
    =-\delta_{m,0}\delta_{n,1}\ffrac{\qm^{-j}}{\qm^{p_+}\!-
      \qm^{-p_+}}
  \end{gathered}
\end{equation}
and then follow the standard step-by-step construction of a Drinfeld
double, based on Eqs.~\eqref{double-def} and~\eqref{double-def-1},
which now become
\begin{gather}\label{commut}
  \begin{gathered}
    \kk\beta=\beta(\kk^{-1}?\kk)\kk, \quad \ep\beta = \beta(?\ep)\one
    + \beta(?\kk^{2p_-})\ep
    -\beta(\ep\kk^{-2p_-}?\kk^{2p_-})\kk^{2p_-},\\
    \fm\beta =\beta(?\fm)\one + \beta(?\kk^{-2p_+})\fm
    -\beta(\fm\kk^{2p_+}?\kk^{-2p_+})\kk^{-2p_+}.
  \end{gathered}
\end{gather}
We here use~\eqref{multiplic}.  The following formulas are then
obtained by direct calculation:
\begin{alignat*}{3}
  \dkk(\kk^{-1}?\kk)&
  =\dkk,&\quad\dkk(?\ep)&=0,&\quad\dkk(?\kk^{2p_-})&=\qp\dkk,\\
  \dkk(\ep\kk^{-2p_-}?\kk^{2p_-})&=0,&\quad\dkk(?\fm)&=0,
  &\quad\dkk(?\kk^{-2p_+})&=\qm^{-1}\dkk,\\
  \dkk(\fm\kk^{2p_+}?\kk^{-2p_+})&=0,&&&&\\
  \fp(\kk^{-1}?\kk)&=\qp^{-1}\fp,&\quad\fp(?\ep)&
  =\ffrac{-\dkk^{2p_-}}{\qp^{p_-}\!- \qp^{-p_-}},&\quad
  \fp(?\kk^{2p_-})&=\fp,\\
  \fp(\ep\kk^{-2p_-}?\kk^{2p_-}) &=\ffrac{-\one}{\qp^{p_-}\!-
    \qp^{-p_-}},&\quad\fp(?\fm)&=0,
  &\quad\fp(?\kk^{-2p_+})&=\fp,\\
  \fp(\fm\kk^{2p_+}?\kk^{-2p_+})&=0,&&&&\\
  \emi(\kk^{-1}?\kk)&=\qm\emi,&\quad\emi(?\ep)&=0,&\quad
  \emi(?\kk^{2p_-})&=\emi,\\
  \emi(\ep\kk^{-2p_-}?\kk^{2p_-})&=0,&\quad\emi(?\fm)
  &=\ffrac{-\dkk^{-2p_+}}{\qm^{p_+}\!- \qm^{-p_+}},
  &\quad\emi(?\kk^{-2p_+})&=\emi,\\
  \emi(\fm\kk^{2p_+}?\kk^{-2p_+}) &=\ffrac{-\one}{\qm^{p_+}\!-
    \qm^{-p_+}},&&&&
\end{alignat*}
Applying the first relation in \eqref{double-def}
to~\eqref{multiplic}, we subsequently obtain
\begin{gather*}
  \coup{\dkk^a}{\pbw_{jmn}}=\q^{aj}\delta_{m0}\delta_{n0},\\
  \coup{\fp^a}{\pbw_{jmn}} =(-1)^a\,\delta_{ma}\delta_{n0}\,
  \ffrac{[a]_+!}{(\qp^{p_-}\!- \qp^{-p_-})^a}
  \qp^{aj+p_-\frac{a(a-1)}{2}},\\
  \coup{\emi^a}{\pbw_{jmn}} =(-1)^a\,\delta_{m0}\delta_{na}\,
  \ffrac{[a]_-!}{(\qm^{p_+}\!- \qm^{-p_+})^a}
  \qm^{-aj+p_+\frac{a(a-1)}{2}},\\
  \coup{\fp^a\emi^b\dkk^c}{\pbw_{jmn}}=\delta_{ma}\delta_{nb}
  \ffrac{(-1)^{a+b}[a]_+![b]_-!}{ {(\qp^{p_-}\!- \qp^{-p_-})}^a
    {(\qm^{p_+}\!- \qm^{-p_+})}^b} \qp^{aj+p_-\frac{a(a-1)}{2}}
  \qm^{-bj+p_+\frac{b(b-1)}{2}}\q^{jc}.
\end{gather*}
It is now straightforward to prove that the $4 p_+^2 p_-^2$ elements
$\{\fp^a\emi^b\dkk^c\}$ with $0\leq a\leq p_+\,{-}\,1$, $0\leq b\leq
p_-\,{-}\,1$, and $0\leq c\leq 4 p_+ p_- -1$ are linearly independent,
cf.~\cite{[FGST]}, and that the relations claimed in the theorem are
indeed satisfied.

\section{Summation over $2p_+ p_-$ consecutive values}
Here, we isolate elementary but bulky formulas needed in the
derivation of modular transformations.  For any $f$ satisfying
$f(r+2p_+p_-)=f(r)$, we have the obvious identity
  \begin{multline*}
    \sum_{r=0}^{2p_+p_- - 1}f(r)
    =\sum_{r'=0}^{p_+ - 1}\sum_{s'=0}^{p_- - 1}f(p_-r'+p_+s')
    +\sum_{r'=0}^{p_+ - 1}\sum_{s'=1}^{p_-}f(p_-r'-p_+s')\\
    =f(0)+f(-p_+p_-)
    + \sum_{r'=1}^{p_+ - 1}\sum_{s'=1}^{p_- - 1}
    \bigl(f(p_-r'+p_+s')+f(p_-r'-p_+s')\bigr)\\
    +\sum_{r'=1}^{p_+ - 1}\bigl(f(p_-r')+f(-p_-r')\bigr)
    +\sum_{s'=1}^{p_- - 1}\bigl(f(p_+s')+f(-p_+s')\bigr).
  \end{multline*}
  
  Next, for $f(r)=g(r)h_\pm(r)$, where $g(r+2p_+p_-)=g(r)$,
  $h_\pm(r+2p_+p_-)=h_\pm(r)$, and in addition $h_\pm(-r)=\pm
  h_\pm(r)$, we have
  \begin{multline}\label{eq:the-previous}
    \sum_{r=0}^{2p_+p_- - 1}
    g(r)h_\pm(r)=g(0)h_\pm(0) + g(-p_+p_-)h_\pm(-p_+p_-)
    \\
    + \sum_{(r',s')\in\setii}\!\!\Bigl(\!\bigl(g(p_-r'+p_+s')
    \pm g(-p_-r'-p_+s')\bigr)h_\pm(p_-r'+p_+s')\\
    +\bigl(g(p_-r'-p_+s')\pm g(-p_-r'+p_+s')\bigr)
    h_\pm(p_-r'-p_+s')\Bigr)\\
    +\sum_{r'=1}^{p_+ - 1}\bigl(g(p_-r')\pm g(-p_-r')\bigr)
    h_\pm(p_-r')
    +\sum_{s'=1}^{p_- - 1}\bigl(g(p_+s')\pm g(-p_+s')\bigr)h_
    \pm(p_-s').
  \end{multline}
  In terms of the combinations
  \begin{align*}
    u^+_\pm(r,s)={}&h_\pm(p_-r+p_+s)+h_\pm(p_-r-p_+s),\\
    u^-_\pm(r,s)={}&h_\pm(p_-r+p_+s)-h_\pm(p_-r-p_+s),
    \quad(r,s)\in\setii,
  \end{align*}
  Eq.~\eqref{eq:the-previous} is written as
  \begin{multline}\label{eq:the-sum}
    \sum_{r=0}^{2p_+p_- - 1}
    g(r)h_\pm(r)
    ={}
    g(0)h_\pm(0) + g(-p_+p_-)h_\pm(-p_+p_-)
    \\
    {}+\half
    \sum_{(r',s')\in\setii}\!\!\Bigl(\!\bigl(g(p_-r'+p_+s')\pm
    g(-p_-r'-p_+s')\\ 
    \shoveright{{}+g(p_-r'-p_+s')\pm
      g(-p_-r'+p_+s')\bigr)u^+_\pm(r',s')\quad}\\
    \kern60pt{}+\bigl(g(p_-r'+p_+s')\pm g(-p_-r'-p_+s')\\
    \shoveright{{}-g(p_-r'-p_+s')\mp
      g(-p_-r'+p_+s')\bigr)u^-_\pm(r',s')\Bigr)}\\
    {}+\sum_{r'=1}^{p_+ - 1}\bigl(g(p_-r')\pm
    g(-p_-r')\bigr)h_\pm(p_-r')
    +\sum_{s'=1}^{p_- - 1}\bigl(g(p_+s')\pm
    g(-p_+s')\bigr)h_\pm(p_-s').
  \end{multline}
This rearrangement of the sum of $2p_+ p_-$-periodic functions over
$2p_+ p_-$ consecutive values is an efficient way to find modular
transformations of the $(p_+,p_-)$-model characters.

\end{document}